\documentclass{./jfm}

\usepackage{graphicx}
\usepackage{epstopdf, epsfig}
\usepackage[dvipsnames]{xcolor}  
\usepackage{hyperref} 
\usepackage{subcaption} 
\usepackage{lipsum}                     
\usepackage{xargs}                      
\usepackage{color} 
\usepackage{comment} 
\usepackage{amsmath} 
\usepackage{cases}
\usepackage[normalem]{ulem}
\usepackage[english]{babel}
\usepackage{blindtext}

\usepackage{pgfplots}
\pgfplotsset{compat=1.10}
\usepgfplotslibrary{fillbetween}
\usetikzlibrary{patterns}

\reversemarginpar

\newcommand{\bigoh}[1]{\mathcal{O}\left( #1 \right)}

\newcommand{\chevron}[1]{\left\langle #1 \right\rangle}

\newcommand{\utau}{u_{\tau}}

\newcommand{\pderiv}[2]{\frac{\partial #1}{\partial #2}} 

\newcommand{\logdutauplus}{\pderiv{\log(\utau)}{X^+}}

\newcommand{\lnubar}{\overline{l_{\nu}}}
\newcommand{\tnubar}{\overline{t_{\nu}}}

\newcommand{\lnuXp}{l_{\nu}(X^+)}
\newcommand{\xp}{x^+}
\newcommand{\Xp}{X^+}
\newcommand{\yp}{y^+}
\newcommand{\zp}{z^+}
\newcommand{\tp}{t^+}
\newcommand{\etap}{\eta^+}

\newcommand{\red}[1]{{\color{red} #1}}

\title{Slow-growth approximation for near-wall patch representation of wall-bounded turbulence}
\author{Sean P. Carney\aff{1} and Robert D. Moser\aff{2,3}\corresp{\email{rmoser@oden.utexas.edu}}}
\shortauthor{S.\ P.\ Carney and R.\ D.\ Moser}
\affiliation{\aff{1} Department of Mathematics, University of California, Los Angeles, CA 90095, USA  
\aff{2} Oden Insititute for Computational Engineering and Sciences, The University of Texas at Austin, TX 78712, USA
\aff{3} Department of Mechanical Engineering, The University of Texas at Austin, TX 78712, USA
}

\begin{document}

\maketitle

\begin{abstract}

Wall-bounded turbulent shear flows are known to 
exhibit universal small-scale dynamics that are modulated by large-scale
flow structures. Strong pressure gradients complicate this characterization, however; 
they can cause significant variation of the mean
flow in the streamwise direction. For such situations, we perform
asymptotic analysis of the Navier-Stokes
equations to inform a model for the effect of mean flow
growth on near-wall turbulence in a
small domain localized to the boundary. The asymptotics are valid
whenever the viscous length scale is small relative to
the length scale over which the mean
flow varies. 
To ensure the correct momentum
environment, a dynamic procedure is introduced that accounts for the
additional sources of mean momentum
flux through the upper domain boundary arising from the asymptotic
terms. Comparisons of the model's
low-order, single-point statistics with those from direct numerical
simulation and well-resolved large eddy simulation of adverse-pressure 
gradient turbulent boundary layers indicate the asymptotic model successfully
accounts for the effect of boundary layer growth on the small-scale
near-wall turbulence.

\end{abstract}

\begin{keywords}
\end{keywords}

\section{Introduction}       

High Reynolds number wall-bounded turbulent shear flows are characterized by a 
separation of scales between the flow in the near-wall region, in which mean 
viscous stresses play an important role, and the flow farther away from the 
wall, where mean viscous effects are negligible. The friction Reynolds 
number $Re_{\tau} = \delta/\delta_{\nu}$ quantifies this separation of scales,
where $\delta$ is the characteristic length scale of the shear layer,
such as a channel half-width, a pipe radius, or a boundary layer thickness, and 
$\delta_{\nu} = \nu/u_{\tau}$ is the viscous length scale, where $\nu$ is the 
kinematic viscosity of the fluid, $u_{\tau} = \sqrt{\tau_w/\rho}$, $\tau_w$ 
is the mean wall shear stress, and $\rho$ is the fluid density. To simulate all
the scales of motion in a wall bounded flow requires $\bigoh{Re_{\tau}^{2.5}}$
and $\bigoh{Re_{\tau}^2}$ spatial degrees of freedom for direct numerical 
simulation (DNS) and large eddy simulation (LES), respectively \citep{Mizuno:2013dc}. 
Even on modern high-performance computing systems, this cost is prohibitively large
for important atmospheric and aeronautical flows, for example, which routinely occur at 
$10^{4} \lesssim Re_{\tau} \lesssim 10^{7}$, \citep{phys_today_turb:2013}. 
Developing reduced order models, such as wall-modeled LES, to overcome
this challenge
requires an understanding of the mutual interactions between small and 
large-scale motions in the outer and near-wall regions. 

Advances in computational power and experimental techniques have enabled a great deal of 
insight into the inner/outer interactions for the canonical zero-pressure
gradient boundary layer and fully developed pipe and channel flows \citep{Smits:2011}. 
It is well established that there is an autonomous near-wall cycle of self sustaining
mechanisms \citep{Moin:1991, Hamilton:1995vu, Jeong:1997uj},  
involving low and high speed streamwise velocity streaks and coherent structures 
of quasi-streamwise vorticity. \citet{Jimenez:1999wf} showed that this cycle of 
near-wall dynamics persists without any input from the turbulence farther away 
from the wall. Large-scale motions, or superstructures, in the
outer layer do indeed impact the near-wall region, however. They modulate the 
turbulent velocity fluctuations and superimpose their energy 
\citep{Hutchins:2007kd, Marusic:2010bb, Ganapathisubramani:2012dh}, and their
influence increases with $Re_{\tau}$ \citep{DeGraaff:2000wm}. 
Spectral analysis of both channel \citep{Lee:2019,wang2021scaling}
and boundary layer flow data \citep{Samie:2018} has demonstrated that, in contrast, 
the dynamics of the small-scale motions in the near-wall region are universal. 
The small-scale, high-wavenumber energy, as well as its production, dissipation, and transport, 
are independent of $Re_{\tau}$.

Based on this characterization of near-wall dynamics, \citet{carney2020} formulated numerical simulations
on near-wall `patch' (NWP) domains whose size scaled in viscous units. Similar to the
numerical experiments of \citet{Moin:1991} and \citet{Jimenez:1999wf}, the 
model used restricted domain sizes and, as in the latter, manipulation of the turbulence 
outside of the near-wall region to simulate \emph{only} the autonomous dynamics 
over the range of scales at which they occur. The model reproduced near-wall
small-scale statistics obtained from direct numerical simulation, confirming that 
the `universal signal' described in \citet{marusic2010,mathis_hutchins_marusic_2011} 
indeed arises from universal dynamics, independent of $Re_{\tau}$ or external 
flow configuration. As a computational model, the near-wall patch (NWP) defined a 
one parameter family of turbulent flows parameterized by the near-wall, viscous-scaled pressure
gradient. Because of its ability to reproduce a variety of well-known features of
high $Re_{\tau}$ wall turbulence at a computational cost that is orders of magnitude 
less than DNS, the model offers a way to efficiently probe the response of near-wall turbulence
 to changes in the mean momentum environment. 

The NWP model was validated against DNS data from channel flows, featuring mild 
favorable-pressure gradients, and a zero-pressure gradient (ZPG) boundary layer. 
Because of their relevance to engineering applications, it is reasonable to ask 
if the NWP model can adequately describe the  near-wall, small-scale dynamics
of flows with adverse-pressure gradients (APGs), especially APG boundary layers.
Although the understanding of scale interactions between inner and outer regions 
in APG boundary layers is less complete than for ZPG flows, 
there has been much progress since the early
experimental studies of \citet{clauser1954turbulent} and \citet{bradshaw1967} and numerical 
simulations of \citet{spalart1987} and \citet{spalart1993experimental}; see also references therein. 
More recent experimental investigations include 
\citet{rahgozar2012}, \citet{harun2013}, \citet{knopp2015}, \citet{knopp2017investigation}, 
\citet{sanmiguel2017,sanmiguel2020} and \citet{romero2022properties}.
Previous large-scale simulations include DNS \citep{na1998direct,skote2002,gungor2016scaling}
and well-resolved LES \citep{hickel2008implicit} of separated boundary flows and 
a separated channel flow \citep{marquillie2008}, while
large scale simulations of 
attached APG boundary layers have been conducted by 
\citet{lee2009structures,kitsios2017,lee2017large,yoon2018} using DNS, as well as by 
\citet{inoue2013adverse,bobke2017,pozuelo2022} using well-resolved LES. 
Simulations over complex airfoil geometries have also been performed with 
both DNS \citep{hosseini2016direct} and well-resolved LES \citep{sato2017large,tanarro2020}.

One observation that has consistently emerged in the literature is that, even when mild,
adverse pressure gradients energize the large-scale structures in both the 
outer layer and the near-wall region of the the boundary layer. 
The increased influence of the large-scales also results in increased modulation effects
on the small-scales \citep{harun2013,lee2017large,yoon2018}, analogous to the
effect of increasing $Re_{\tau}$ in ZPG boundary layers. Although APGs have been 
shown to energize the small-scale motions in the outer region of a boundary layer \citep{sanmiguel2020}, 
less appears to be known about the small-scale energy in the near-wall region. 
After filtering out contributions from spanwise wavelengths $\lambda_z/\delta_{\nu} \gtrsim 180$,
\citet{lee2017large} found the small-scale contribution to both the streamwise 
velocity variance and the Reynolds shear stress increased with pressure gradient, while 
\citet{sanmiguel2020} found that the small-scale contributions to the streamwise velocity 
variance from motions with streamwise wavelengths $\lambda_x/\delta_{\nu} \lesssim 4300$ 
was independent of the pressure gradient strength. 
One objective of the current work is to use a near-wall patch computational model to investigate 
the extent to which the small-scale, near-wall dynamics are responsible 
for the low order flow statistics of adverse-pressure gradient flows observed in 
experiments and large-scale simulations. 
Since the NWP model simulates \emph{only} the small-scales motions, isolated from 
large-scale influences, any differences between its statistical profiles 
and those from DNS or large-scale simulations can reasonably be attributed
to the superposition and modulation effects of the large-scale motions that are missing. 
In this way, the model is similar to the computational 
`experiments' of \citet{Moin:1991} and \citet{Jimenez:1999wf} in which near-wall turbulence 
is artificially manipulated and compared to a `real', unmodified flow. 

The near-wall patch model previously developed in \citet{carney2020} can 
be considered the lowest order 
asymptotic description of small-scale, near-wall dynamics in which the mean pressure
gradient, the momentum flux from the outer flow, and the mean wall shear stress are
all uniform in time and space on the scale of the computational domain. 
To account for the relatively rapid downstream development of mean quantities 
in adverse-pressure gradient boundary layers \citep{kitsios2017}, we develop in the 
present work a higher-order approximation that allows the mean wall shear stress to 
develop slowly in the streamwise direction. Similar to \citet{Spalart:1988,guarini2000,maeder2001}
and \citet{topalian2017}, asymptotic analysis is used to derive a set of `homogenized' equations 
that describe the mean effect of streamwise development on the near-wall dynamics. 

If such a higher order computational model can be shown to accurately reproduce the 
near-wall, small-scale features of adverse-pressure gradient flows, 
or, more generally, for flows that feature asymptotic growth of mean quantities 
in the near-wall region, it could be used to inform a pressure-gradient dependent 
wall model for LES \cite{Piomelli:2002,Bose:2018}. In this setting, the model is a pressure-gradient-dependent
analogue to the experimentally determined `universal signal' of \citet{mathis_hutchins_marusic_2011}. 
Additionally, the model could be used to study the interaction 
between small-scale  near-wall turbulent dynamics and more complicated physical 
processes such as heat transfer, chemical reactions, turbophoresis, or surface 
roughness. 

The rest of the paper is organized 
as follows: $\S$\ref{sec:motivation} motivates the slow-growth near-wall patch model 
and contains the multiscale 
asymptotic analysis on which the model is based. 
Section \ref{sec:formulation} then details 
 the computational model and the numerical method used to integrate
the equations of motion. Section \ref{sec:numerical_results} provides a comparison 
between the statistics generated by the model and the corresponding 
quantities from DNS for the cases of both zero and mild adverse pressure gradients. 
It is followed by a discussion and conclusions in sections \ref{sec:discussion} and \ref{sec:conclusions}, 
respectively.

\subsection{Mathematical notation and nonmenclature}
In the following discussion, the velocity components in the
streamwise ($x$), wall-normal ($y$) and spanwise ($z$) directions
are denoted as $u$, $v$, and $w$, respectively, and when
using index notation, these directions are labeled 1, 2, and
3, respectively. 
 The expected value is denoted with 
angle brackets (as in $\langle \cdot \rangle$), and upper
case $U$ and $P$ indicate the mean velocity and pressure, so 
that $\langle u_i \rangle = U_i$. The velocity and pressure 
fluctuations are indicated with primes, e.g. $u_i = U_i + 
u_i'$. Partial derivatives shortened to $\partial_i$ signify
$\partial/\partial x_i$, differentiation in the direction $x_i$.
 The mean advective derivative is
$D(\cdot)/Dt = \partial_t (\cdot) + U_j \partial_{j} (\cdot)$, where
Einstein summation notation is implied. 
In general, repeated indices
imply summation, with the exception of repeated Greek indices. 
Lastly, 
the superscript `$+$' denotes non-dimensionalisation with the 
kinematic viscosity $\nu$ and the friction velocity $u_{\tau}$.

\section{Motivation}\label{sec:motivation}       
\subsection{Fundamental modeling assumptions}\label{subsec:scale_sep_assuptions}
Intrinsic to the computational model is the assumption of a
  separation of temporal and spatial scales between the small-scale
  turbulence arising from the autonomous near-wall dynamics and the
  large-scale outer-layer turbulence; this separation of scales
    occurs when the friction Reynolds number of the flow is asymptotically large.
The near-wall dynamics 
are thus considered to be in local equilibrium with both the pressure 
gradient and momentum flux environment in which they evolve, as explored
in \citet{Zhang:2016,Chernyshenko:2021}.
In the previous near-wall patch formulation \citep{carney2020}, 
it was further assumed these quantities were uniform in space and
time on the scale of the dynamics being simulated. 
In the current work, the assumption of a constant 
pressure gradient is retained, however, the local mean wall shear stress 
is allowed to vary slowly in the streamwise direction, 
which should allow for a higher order asymptotic description 
of near-wall turbulence than before. 
In particular, it is assumed that the rate of change of the viscous
length scale is asymptotically small.
Under these assumptions, the near-wall model will be representative of a variety 
of flows that are not in equilibrium overall, including those
with non-constant pressure gradients.  However, the modeling approach breaks
down, for example, for a boundary layer near separation; see  
section \ref{sec:discussion} for some remarks on this case.

\subsection{Growth effects in the near-wall region of turbulent boundary layers}
\label{subsec:apg_tbl_discussion}
Consider a flat plate turbulent boundary layer that is homogeneous in the 
spanwise direction and under the influence of 
a pressure gradient in the streamwise direction. Let the pressure gradient 
be parameterized by 
\begin{equation}\label{eq:Clauser}
\beta = \frac{\delta^{\ast}}{\tau_w} \frac{dP_{\infty}}{dx},
\end{equation}
which is the standard nondimensional Clauser parameter, where $\delta^{\ast}$ is the
boundary layer displacement thickness, $\tau_w$ is the mean shear stress at the wall, and 
$dP_{\infty}/dx$ is the far-field pressure gradient with a unit density. 
For any $\beta\in [0,\infty)$ the boundary layer as a whole will grow in the streamwise
direction; a fortiori, so too will the near-wall region. 
Below we make two observations about the effect this growth has on the near-wall region of 
turbulent boundary layers (TBLs). These observations motivate the 
multiscale asymptotic analysis that will inform the slow-growth near-wall 
patch model.  

\begin{figure}
  \begin{center}
	\includegraphics[width=1.0\textwidth]{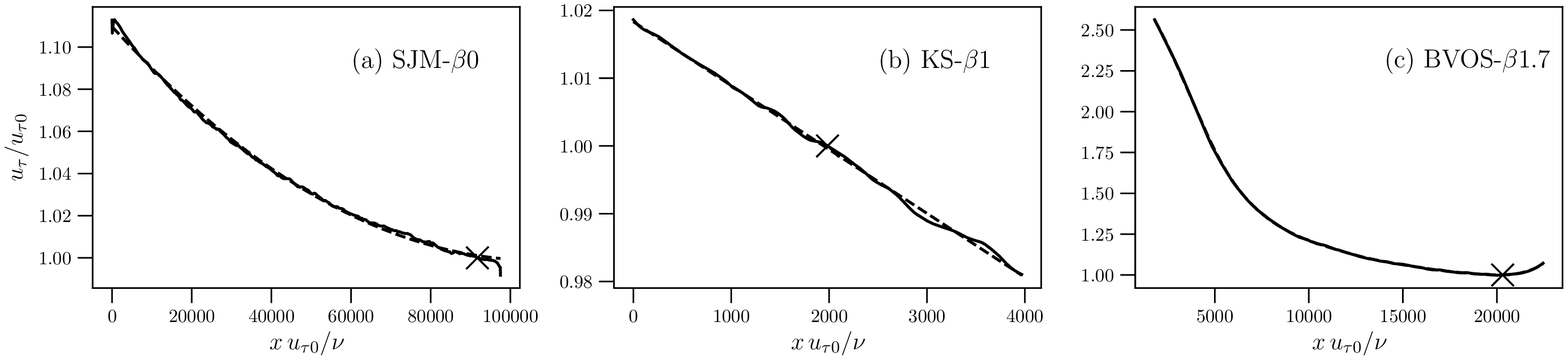}
  \end{center}
\caption{
Friction velocity $u_{\tau}/u_{\tau 0}$ versus streamwise location $x\, u_{\tau 0}/\nu$
for each case in table \ref{table:dns_eps}, where $u_{\tau 0}$ is the value 
of the friction velocity at the locations marked `$\times$'. 
The dashed lines 
in (a) and (b) show quadratic and linear least-squares approximations, respectively, 
while in (c) they show a piecewise cubic least-squares fit.     
}  \label{fig:utauvsx}
\end{figure}

The first observation is that the near-wall region of TBLs grows more rapidly with increasing $\beta$. 
As $\tau_w$ evolves downstream, so too does the viscous length scale 
characterizing the local near-wall scaling. Figure \ref{fig:utauvsx} illustrates 
the streamwise evolution of the friction velocity $u_{\tau}$ for the three large-scale
simulation
cases considered throughout this work, namely SJM-$\beta0$ \citep{sillero:2013}, KS-$\beta1$
\citep{kitsios2017} and BVOS-$\beta1.7$ \citep{bobke2017}, labeled by the value of 
the Clauser parameter \eqref{eq:Clauser} at the streamwise locations 
marked in each case by `$\times$' in the figure. Each friction velocity and streamwise 
location is scaled by the kinematic viscosity and value of $u_{\tau}$ at these particular
locations, denoted below by $\overline{x}$. 

As described
  in section \ref{sec:formulation}
below, the model statistics are a function of the wall-normal
direction only;
that is, statistics
are homogeneous in the stream and spanwise direction. In constrast, statistics of the 
TBLs to which the model is compared are only homogeneous in the spanwise direction. Hence, 
comparisons can only be made
at particular streamwise locations $\overline{x}$. 
For SJM-$\beta0$, $\overline{x}$ is selected as
the location with the largest value of $Re_{\tau}$ for which statistical profiles are reported, 
while $\overline{x}$ is selected for the BVOS-$\beta1.7$ case to maximize $Re_{\tau}$ before 
boundary effects from the `fringe region' used to periodically match the TBL inlet and 
outlet profiles \citep{bobke2017} affect the statistics (this fringe region corresponds to 
the locations in figure \ref{fig:utauvsx}(c)
where $\partial u_{\tau}/\partial x$ is positive). 
The reason for maximizing $Re_{\tau}$ is to make comparisons at locations where the 
modeling ansantz just described in section \ref{subsec:scale_sep_assuptions} is most valid.
Since both $\beta$ and $Re_{\tau}$ are approximately 
constant throughout the domain for the KS-$\beta1$ simulation, $\overline{x}$ is simply 
taken in the middle.


The rate of change of
the friction velocity with respect to streamwise position $x$ 
can be used to define a length scale $L$ that quantifies the streamwise distance over which 
the near-wall region grows. Define $L$ by 
\begin{equation}\label{eq:L_defn}
L^{-1} = \frac{1}{u_{\tau}} \frac{\partial u_{\tau}}{\partial x} 
\end{equation}
where both $u_{\tau}$ and $\partial u_{\tau}/\partial x$ are evaluated at $\overline{x}$. 
Using also the viscous length scale $l_{\nu}$ at $\overline{x}$, define the nondimensional 
asymptotic parameter 
\begin{equation}\label{eq:eps_defn}
\epsilon = \frac{l_{\nu}}{L} = \frac{ \nu }{u_{\tau}^2}  \frac{\partial u_{\tau}}{\partial x}.
\end{equation}
The value of $\epsilon$ for each TBL case shown in figure \ref{fig:utauvsx} ranges from
approximately $10^{-5}$ to $10^{-7}$, as listed in table \ref{table:dns_eps}. For SJM-$\beta0$
and KS-$\beta1$, the derivative $\partial u_{\tau}/\partial x$ is estimated by differentiating 
a quadratic and linear least-squares approximation to $u_{\tau}$, respectively. 
The $u_{\tau}$ data is relatively noisy in the BVOS case, so the data is first filtered
with a Savitzky-Golay filter \citep{savgol:1964}, and it is then separately fit to a piecewise 
cubic spline interpolant. The derivative $\partial u_{\tau}/\partial x$ is then taken to be the
average of the derivatives of these two approximations. 
\begin{table}
  \begin{center}
  \def~{\hphantom{0}}
    \begin{tabular}{c c c c c c}
     Large scale simulation & $Re_{\tau}$ & $\beta$  & $dP^+/dx^+$  & $\epsilon$   \\
\hline
     SJM2000   & 1989.4  & 0    & $0$~       &  $-2.985\cdot10^{-7}$~   \\
     KS-$\beta$1   & 184.6 & 1.02   & $5.503\cdot 10^{-3}$~      &  $-9.4303\cdot 10^{-6}$~   \\
     BVOS-$\beta$1.7 & 760.1  & 1.72   & $8.981 \cdot 10^{-3}$~          &  $-1.064\cdot 10^{-5}$~   \\
    \end{tabular}
    \caption{Parameters from the large-scale simulations considered at the
streamwise location marked `$\times$' in figure \ref{fig:utauvsx}. 
}
  \label{table:dns_eps}
  \end{center}
\end{table}

The dimensionless parameter $\epsilon$ is readily seen to be the (negative) streamwise rate of change 
of the viscous length scale, and it can also be taken as the inverse Reynolds number based on 
$L$ and $u_{\tau}$:
\begin{equation}
Re_{\epsilon} = L/l_{\nu}. \nonumber
\end{equation}
The asymptotic analysis detailed in Section \ref{subsec:multiscale_analysis} is then 
valid for asymptotically large $Re_{\epsilon}$; 
the infinite $Re_{\epsilon}$ limit corresponds to zero-growth of the near-wall layer, 
e.g.\ in a channel or pipe flow, while the vanishing $Re_{\epsilon}$ limit corresponds to 
boundary layer separation. 

\begin{figure}
  \begin{center}
	\includegraphics[width=1.0\textwidth]{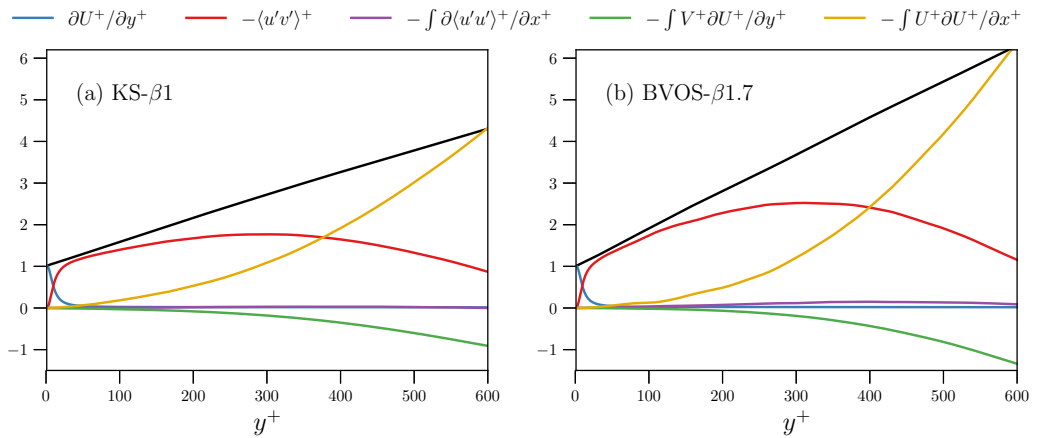}
  \end{center}
\caption{Contributions to the total stress \eqref{eq:tot_stress_apg} versus $y^+$ for the KS-$\beta1$ case (a) 
and BVOS-$\beta1.7$ case (b). The black curves show the sum of all contributions, i.e.\ the right-hand side of 
\eqref{eq:tot_stress_apg}.
} 
\label{fig:tbl_stress_balances}
\end{figure}

The second observation about growth effects in the near-wall region of TBLs
is that there is an increase in momentum flux towards the wall with increasing $\beta$; in 
particular, the Reynolds shear stress increases in magnitude. To quantify this, 
consider the mean stress balance in viscous units for a TBL with (locally) constant
pressure gradient $dP/dx$:
\begin{align}
1 + \frac{dP^+}{dx^+} y^+ =&  \pderiv{U^+}{y^+} -  \chevron{u'v'}^+ \nonumber \\
&-  \int_0^{y^+} \left( U^+ \pderiv{U^+}{x^+} + V^+ \pderiv{U^+}{s^+} 
+ \pderiv{}{x^+}\chevron{u'u'}^+ 
- \frac{\partial^2 U^+}{\partial x^+ \partial x^+} \right) ds.  \label{eq:tot_stress_apg}
\end{align}
Note that here $s$ is just a dummy variable of integration. As $dP^+/dx^+$ increases, so too 
must the mean total stress on the right hand side of \eqref{eq:tot_stress_apg}. 
Figure \ref{fig:tbl_stress_balances} illustrates 
this balance for the two mild adverse-pressure-gradient TBL flows KS-$\beta1$ and BVOS-$\beta1.7$,
where, as before, quantities are scaled in viscous units at  
the locations marked `$\times$' in figure \ref{fig:utauvsx}. 
Besides the Reynolds shear stress, the mean convective terms make a significant 
contribution to the stress balance, even in the near-wall region $y^+ \le 300.$
In contrast, the mean convective terms from SJM-$\beta0$ make a negligible
contribution to the overall stress balance (not shown). 
In all cases, the mean viscous and turbulent fluctuation growth terms in \eqref{eq:tot_stress_apg}
do not make a meaningful contribution to the total stress balance.

Large-scale numerical simulation of turbulent boundary layers properly account
for the effect of boundary layer growth on the near-wall region simply 
by computing on domains that are sufficiently large. Clever `recycling' and rescaling 
techniques \cite{lund1998,Colonius:2004,araya2011,sillero:2013} are typically used to increase 
size of the part of the domain containing ``healthy'' turbulence (i.e.\ turbulence not 
impacted by inflow-outflow artifacts) while ensuring 
the simulations remain computationally affordable.  
\citet{Spalart:1988}, however, took an alternative approach to achieve this goal. 
Assuming a scale separation between the size of the boundary layer and the streamwise 
length over which it develops, asymptotic analysis was used to determine a 
set of `homogenized' equations of motion featuring the standard Navier-Stokes equations 
augmented with additional terms modeling the effect of boundary layer growth. 

Inspired by the approach in \citet{Spalart:1988}, we now describe a multiscale analysis 
to build a near-wall patch representation of the near-wall, small-scale
dynamics of turbulent flows with asymptotically small $\epsilon$. 
In this case the separation of scales assumed in \citet{Spalart:1988} is 
expected to be even stronger, since only the near-wall layer is of 
interest, in contrast to the entire boundary layer.

\subsection{Multiscale asymptotic analysis}\label{subsec:multiscale_analysis}
The goal of the following analysis is to derive a set of equations 
for a near-wall patch domain that can produce accurate near-wall 
statistics for spatially developing flows. 

First, if the viscous length scale evolves over
distances that are asymptotically large relative to its local values, it is sensible to
hypothesize a scaling relationship for the fluid velocity of the form 
\begin{equation}\label{eq:scaling_assumption}
u_i(x,y,z) = u_{\tau}(\epsilon x^+)u_i^+(x^+, y^+, z^+) ,
\end{equation}
where $\epsilon$ is the dimensionless order parameter \eqref{eq:eps_defn} 
and $u_i^+$ is considered statistically homogeneous in the stream and spanwise directions. 
This homogeneity will allow for the use of periodic boundary conditions (and Fourier spectral 
discretizations) for the near-wall patch domain, as in \citet{Spalart:1988}.
The superscript `$+$' here denotes nondimensionalization by the local viscous 
scale. Equation \eqref{eq:scaling_assumption} is nothing but the standard near-wall viscous 
scaling where the friction velocity evolves slowly in the streamwise direction.

For some specific streamwise
location $\overline{x}$, let 
$$
\lnubar := l_{\nu} \Big\vert_{x = \overline{x}}
$$
denote the local viscous length scale, and let $L$ be the length scale defined by 
\eqref{eq:L_defn}; that is, the inverse of the logarithmic derivative of $u_{\tau}$.
For some given $\epsilon$, define $X = \epsilon x$, as well as the new 
coordinates
\begin{equation}\label{eq:coord_trans}
(x,y,z) \mapsto (x/\lnubar,\,\, y\, \utau(X^+)/\nu,\,\, z/\lnubar) =: (\xp,\etap,\zp).
\end{equation} 
Note that in the definition of $\eta^+$, the argument in the friction 
velocity $u_{\tau}$ is $X^+ = \epsilon x/\lnubar = x/L$. 

The plan now is to first transform the 
incompressible Navier-Stokes equations from Cartesian
 to $(x^+, \eta^+, z^+)$
coordinates and then insert the scaling hypothesis 
\eqref{eq:scaling_assumption} into the result. 

To transfrom the mass and momentum equations to the new coordinates
\eqref{eq:coord_trans}, first note that derivatives transform as
\begin{align}
\pderiv{}{x} &\mapsto 1/\lnubar \,\,\pderiv{}{\xp} +  \epsilon/\lnubar\,\, \etap \logdutauplus \pderiv{}{\etap} \nonumber \\
\pderiv{}{y} &\mapsto 1/l_{\nu}(\Xp) \,\, \pderiv{}{\etap} \nonumber \\
\pderiv{}{z} &\mapsto 1/\lnubar \,\,\pderiv{}{\zp}, \nonumber 
\end{align}
where $\lnuXp = \nu/\utau(\Xp)$. Inserting the transformations in 
the continuity equation 
\begin{equation}
\pderiv{u_i}{x_i}  = 0  \nonumber
\end{equation} 
gives
\begin{equation} \label{eq:continuity_sg_coord}
\pderiv{u}{\xp}  + \lnubar/\lnuXp \pderiv{v}{\etap}
 + \pderiv{w}{\zp} + \epsilon \, \etap\,\logdutauplus \pderiv{u}{\etap} =0. 
\end{equation}
After additionally scaling by the viscous time scale at $\overline{x}$
$$ 
\overline{t_{\nu}} := \frac{\nu}{u_{\tau}^2} \Big\vert_{x = \overline{x}}
$$
the streamwise component of the momentum equation transforms to 
\begin{align}
\lnubar/\tnubar \, \pderiv{u}{\tp} + u \pderiv{u}{\xp} + \lnubar/l_{\nu}(\Xp) v \pderiv{u}{\etap} 
+ w \pderiv{u}{\zp} + \epsilon \, \etap\, \logdutauplus u \pderiv{u}{\etap}& \nonumber\\
+ \pderiv{p}{\xp} + \epsilon\, \etap\,\logdutauplus \pderiv{p}{\etap}- \nu/\lnubar \frac{\partial^2 u}{\partial\xp\partial \xp}
- \nu/\lnubar \frac{\partial^2 u}{\partial\zp\partial\zp} 
& \nonumber \\
- \nu \, \lnubar/l_{\nu}^2(X^+) \frac{\partial^2 u}{\partial \eta^+ \partial \eta^+} 
- 2\epsilon\nu \, \etap /l_{\nu}(\Xp) \, \logdutauplus \frac{\partial^2 u}{\partial \xp \partial \etap}& = 0, \label{eq:mom_pre_freeze}
\end{align}
where the $\bigoh{\epsilon^2}$ terms have been dropped. Similar terms
appear for the other components. 
So far, the equations have simply been recast into new coordinates. 
In \eqref{eq:mom_pre_freeze}, the advective derivative is on the first 
line while the pressure gradient is on the second; the viscous 
terms are on both the second and third. 

The next step is to hypothesize that the velocity and pressure
fields scale with $u_{\tau}(X^+)$ and $u_{\tau}^2(X^+)$,
respectively, as in \eqref{eq:scaling_assumption}. Using the 
superscript `$+$' to denote this scaling, 
the continuity equation \eqref{eq:continuity_sg_coord} becomes
\begin{equation}\label{eq:continuity_post_ansatz}
\pderiv{u^+}{\xp} + \lnubar/\lnuXp \pderiv{v^+}{\etap} + \pderiv{w^+}{\zp}
+ \epsilon \, \logdutauplus \pderiv{}{\etap}\left(\etap u^+ \right) = 0. 
\end{equation}
Recall that the multiscale assumption underlying this analysis 
asserts that, at any given streamwise location, the $\epsilon$-dependent 
slow-growth terms evolve over asymptotically large distances relative
to the local viscous length scale; in particular then at $x = \overline{x}$, 
\eqref{eq:continuity_post_ansatz} simplifies to 
\begin{equation}\label{eq:continuity_sg_full}
\left(\pderiv{u^+}{\xp} + \pderiv{v^+}{\yp} + \pderiv{w^+}{\zp}\right)
+ \epsilon \, \pderiv{}{\yp}\left(\yp u^+ \right) = 0, 
\end{equation}
since at $x = \overline{x}$
\begin{equation*}
\logdutauplus = \frac{\overline{l_{\nu}}}{\epsilon} \pderiv{\log(\utau)}{x}\, (\overline{x}) =1
\end{equation*}
and $l_{\nu} (X^+) = \lnubar$. Note that in \eqref{eq:continuity_sg_full}, $y^+$ denotes $\eta^+$ at $\overline{x}$. 
The same procedure of inserting the scaling assumptions and insisting 
they hold at $x = \overline{x}$ results in 
\begin{align}
\pderiv{u^+_i}{\tp} +  u^+ \pderiv{u^+_i}{\xp} + v^+ \pderiv{u^+_i}{\yp} 
+  w^+ \pderiv{u^+_i}{\zp} +  \epsilon\, u^+ \pderiv{}{\yp}\left(\yp u_i^+\right) & \nonumber\\
+ \left(\pderiv{p^+}{\xp} + \epsilon\,\left( \yp\, \pderiv{p^+}{\yp} + 2 p^+\right)\right)\delta_{1i} 
+ \pderiv{p^+}{\yp} \delta_{2i} + \pderiv{p^+}{\zp} \delta_{3i} & \nonumber\\
- \left( \frac{\partial^2 }{(\partial\xp)^2} + \frac{\partial^2 }{(\partial\yp)^2} +  \frac{\partial^2 }{(\partial\zp)^2}\right) u_i^+  
-2 \epsilon \frac{\partial^2}{\partial \xp \partial \yp}\left( \yp u^+_i \right) & = 0 \label{eq:mom_post_freeze} 
\end{align}
for the $i$-th component of the momentum equation; again the $\bigoh{\epsilon^2}$
have been neglected. 

 Equation \eqref{eq:mom_post_freeze} contains $\bigoh{\epsilon}$ terms 
originating from convective, pressure, and viscous effects. Recall, however, that 
the contribution to the mean stress balance from the viscous streamwise growth
term (the final term in the integral in \eqref{eq:tot_stress_apg}) is 
negligible in the adverse-pressure gradient flows discussed in Section 
\ref{subsec:apg_tbl_discussion}. Hence, the $\bigoh{\epsilon}$ viscous terms are dropped, 
as was done in \citet{Spalart:1988}. Similarly, the $\bigoh{\epsilon}$ 
pressure terms are dropped, since the pressure gradient is assumed to 
be constant over the length scales of the near-wall patch domain (as mentioned in  
section \ref{subsec:scale_sep_assuptions}).   
Thus, only the convective growth terms remain. 


Using index notation, the simplified momentum equation becomes
\begin{equation}\label{eq:mom_post_discard}
\pderiv{u_i^+}{t^+} + u_j^+ \pderiv{u_i^+}{x_j^+} + \pderiv{p^+}{x_i^+} 
- \frac{\partial^2 u_i^+}{\partial x_j^+ \partial x_j^+} 
+ \epsilon u^+ \pderiv{}{y^+} \big( y^+ u_i^+\big)  = 0. 
\end{equation}
For numerical purposes, it is useful to rewrite \eqref{eq:mom_post_discard} in conservative
form. Because of the slow-growth contribution to the continuity equation 
\eqref{eq:continuity_sg_full}, however, 
an additional $\bigoh{\epsilon}$ convective term appears:
\begin{equation}\label{eq:mom_post_discard_conservative}
\pderiv{u_i^+}{t^+} + \pderiv{}{x_j^+} \big( u_i^+ u_j^+\big) + \pderiv{p^+}{x_i^+} 
- \frac{\partial^2 u_i^+}{\partial x_j^+ \partial x_j^+} 
+ \epsilon \left( u^+ u_i^+  + \pderiv{}{y^+} \big( y^+ u^+ u_i^+\big) \right) = 0. 
\end{equation}
Note that this multiscale analysis was carried out starting with the incompressible 
Navier-Stokes equations in Cartesian coordinates written in convective form. If 
instead one starts with the equations written in conservative form, makes the 
same coordinate transformation \eqref{eq:coord_trans} and scaling assuming 
\eqref{eq:scaling_assumption}, and retains
only the $\bigoh{\epsilon}$ convective terms, then equation 
\eqref{eq:mom_post_discard_conservative} will result. 


\subsection{A priori test of SG model}
From the scaling assumption \eqref{eq:scaling_assumption} the velocity
components $u_i^+$ are homogeneous in the stream and spanwise directions. 
At statistical equilibrium, the slow-growth continuity equation 
\eqref{eq:continuity_sg_full} then implies that
\begin{equation}\label{eq:apriori_sg}
V^+ = -\epsilon y^+ U^+. 
\end{equation}
Using data from the adverse-pressure-gradient simulations KS-$\beta1$ and BVOS-$\beta1.7$, 
one can use \eqref{eq:apriori_sg} as an \emph{a priori} test
of the slow-growth asymptotics just detailed. 
In general, the relationship \eqref{eq:apriori_sg} is expected to be relatively accurate
close to the wall
and at large $Re_{\tau}$ and large $Re_{\epsilon}$, since 
it was derived under `slowly developing' viscous scaling ansatz \eqref{eq:scaling_assumption}. 

\begin{figure}
  \begin{center}
	\includegraphics[width=1.0\textwidth]{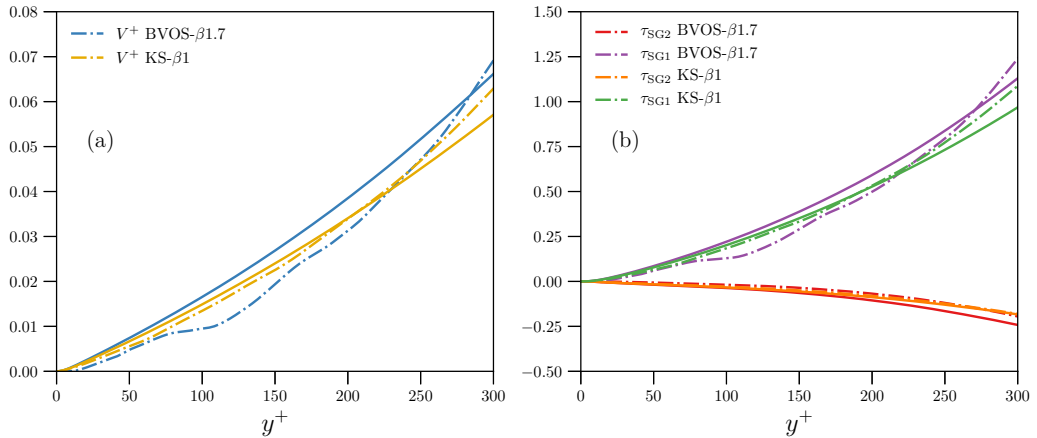}
  \end{center}
\caption{
A priori slow-growth approximation of (a) the mean wall-normal velocity $V^+$ 
and (b) the stress terms \eqref{eq:sg_apriori1} and \eqref{eq:sg_apriori2} 
for KS-$\beta1$ and BVOS-$\beta1.7$. The approximations are shown in solid 
lines, while the quantities taken directly from the large-scale simulations are shown 
in dash-dotted lines. 
} 
\label{fig:apriori_sg}
\end{figure}

Figure \ref{fig:apriori_sg}(a) illustrates that the expression on the right-hand side of 
\eqref{eq:apriori_sg} matches the true wall-normal mean velocity $V^+$ up to a relative 
error of at most $8.6\%$ for the KS-$\beta1$ case. The accuracy is not as high in the BVOS-$\beta1.7$
case; however, the $V^+$ profile is relatively noisy in this case. Since the NWP model 
detailed below aims to simulate wall turbulence in the region $y^+ \in [0,300]$, the profiles 
are shown in this range. In both cases, the discrepancy between the wall-normal mean velocity and its 
approximation increases for
$y^+ \gtrsim 300$, since, at the relatively low values of $Re_{\tau}$ 
at which the large-scale simulations were conducted, wall scaling
becomes inappropriate at relatively low values of $y^+$.

Slow-growth approximations can also be evaluated a priori for the 
mean convection terms which were shown in Figure \ref{fig:tbl_stress_balances} to be 
important to the stress balance \eqref{eq:tot_stress_apg} of adverse-pressure-gradient
turbulent boundary layers in 
the near-wall region. Equation \eqref{eq:apriori_sg} implies that, in 
viscous units, 
$-\int_0^y V \partial_y U \, ds$ is approximated by 
\begin{equation} \label{eq:sg_apriori1}
\tau_{\rm SG 1}^+ := \int_0^{y^+}\epsilon \, s^+ U^+ \pderiv{U^+}{s^+} \, ds^+,
\end{equation}
while the coordinate change \eqref{eq:coord_trans} and scaling assumption 
\eqref{eq:scaling_assumption} imply that in viscous units, $\int_0^y U \partial_x U \, ds$ 
is approximated by 
\begin{equation}\label{eq:sg_apriori2}
\tau_{\rm SG 2}^+ := \int_0^{y^+} \epsilon\, U^+ \pderiv{}{s^+}(s^+ U^+) \, ds^+.
\end{equation}
Figure \ref{fig:apriori_sg}(b) shows that \eqref{eq:sg_apriori1} and \eqref{eq:sg_apriori2} 
are indeed accurate a priori approximation of the 
mean convection terms in the stress balance for the near-wall region $y^+ \le 300$. The 
relative errors are no larger than $9\%$; the exception is approximation \eqref{eq:sg_apriori1}
for the BVOS-$\beta1.7$ case. It inherits the noise from the $V^+$ profile and hence is not as accurate.

\section{Model Formulation}\label{sec:formulation}       
\subsection{Mathematical formulation}
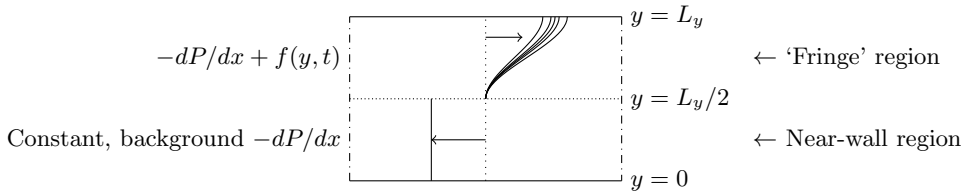
\begin{figure}                                
\centering
\begin{tikzpicture}[xscale=5.4,yscale=5.4]
\draw[dashdotted](0,0) -- (0,0.4);
\draw (0,0.4) -- (.666,0.4);                           
\draw[dashdotted] (.666,0.4) -- (.666,0.0);
\draw[black] (0,0.0) -- (.666,0.0);
\draw[densely dotted] (0,0.2) -- (.666,0.2);

\draw[variable=\y, domain=0:0.2] plot({0.20},{\y});  
\draw[->] (0.333,0.1) -- (0.2,0.1);

\draw[variable=\y, domain=0.2:0.4] plot({-50*\y*\y*\y + 45*\y*\y - 12.00*\y + 1.333},{\y});   
\draw[variable=\y, domain=0.2:0.4] plot({-42.5*\y*\y*\y + 38.25*\y*\y - 10.2*\y + 1.183},{\y});
\draw[variable=\y, domain=0.2:0.4] plot({-35*\y*\y*\y + 31.5*\y*\y - 8.4*\y + 1.033},{\y});   
\draw[variable=\y, domain=0.2:0.4] plot({-40*\y*\y*\y + 36*\y*\y - 9.6*\y + 1.133},{\y});   
\draw[variable=\y, domain=0.2:0.4] plot({-45*\y*\y*\y + 40.5*\y*\y - 10.8*\y + 1.233},{\y});   
\draw[->] (0.333,0.35) -- (0.423,0.35);

\draw[dotted] (0.333,0.0) -- (.333,0.4);

\draw (0., 0.1)node[left]{Constant, background $-dP/dx$} ;
\draw (0.966, 0.1)node[right]{$\leftarrow$ Near-wall region} ;
\draw (0., 0.3)node[left]{$-dP/dx + f(y,t)$} ;
\draw (0.966, 0.3)node[right]{$\leftarrow$ `Fringe' region} ;
\draw (.666,0)node[right]{$y=0$};
\draw (.6660,0.2)node[right]{$y = L_y/2$};
\draw (.6660,0.4)node[right]{$y = L_y$};
\end{tikzpicture}
\caption{
The fluid is subject to periodic boundary conditions at the 
(dash-dotted) side walls, homogeneous Dirichlet/Neumann conditions
at the upper boundary $y=L_y$, and the no-slip condition 
at the wall $y=0$. In addition to the constant pressure gradient assumed to be present in the 
near-wall layer, the model includes a time-dependent, auxiliary forcing 
function $f$ (depicted here at multiple realizations in time) in a 
``fringe region" $L_y/2 \le y \le L_y$ to make up for the momentum not transported at the computational boundary $y=L_y$.
}
\label{fig:fringe_region} 
\end{figure}

The goal of the computational model is to simulate the  
small-scale turbulent dynamics in the near-wall region as a function of an imposed
gradient only in a small, rectangular domain 
$\Omega = [0,L_x]\times[0,L_y]\times[0,L_z]$ localized to the 
boundary. Besides the physical wall at $y=0$ where the no-slip condition is applied, 
the other computational boundaries are non-physical and located where, 
in a large-scale simulation, there is a region of chaotic 
nonlinear dynamics. At the sidewalls, periodic boundary conditions
are used. This is consistent with the main scaling assumption \eqref{eq:scaling_assumption}
underlying the multiscale analysis in section \ref{subsec:multiscale_analysis}, since 
the velocity fields evolved in time are assumed to be statistically homogeneous 
in the stream and spanwise directions. Any statistical inhomogeneities are modeled
by $\bigoh{\epsilon}$ slow-growth terms. At the upper computational boundary $y=L_y$, 
homogeneous Neumann and Dirichlet conditions are prescribed for the stream/spanwise
and wall-normal velocities, respectively. Since these conditions do not allow for any 
momentum flux through the computational boundary, the model includes a `fringe
region' $y \in [L_y/2, L_y]$ in which the flow is forced to provide 
the momentum that is transported into the near-wall region (see figure \ref{fig:fringe_region}
for an illustration). The forcing $f$ is non-zero only 
in the fringe region, and, given a constant pressure 
gradient $dP/dx$, it injects momentum in such a way that ensures that the model's  
wall shear stress at statistical equilibrium is unity, as in \citet{carney2020}. 
Because of the $\bigoh{\epsilon}$ slow-growth contributions to the momentum equation 
derived in section \ref{subsec:multiscale_analysis}, $f$ is time-dependent; the
precise details are given after introducing the model equations of motion below.  

The model equations of motion posed on the domain 
$\Omega$ are
based on the slow-growth continuity \eqref{eq:continuity_sg_full}
and momentum equations and \eqref{eq:mom_post_discard_conservative} 
from the asymptotic analysis 
of Section \ref{subsec:multiscale_analysis}, 
and they are discretized in the statistically homogeneous 
stream and spanwise directions with a Fourier-Galerkin method; 
however, there are a few differences in the equations that govern 
the horizontal averages 
\begin{equation}
\overline{u}_i(y,t) = \frac{1}{L_x}\frac{1}{L_z} \int_0^{L_z} \int_0^{L_x} u_i(x,y,z,t) \, dx \, dz \nonumber
\end{equation}
and the fluctuations $u_i' = u_i - \overline{u}_i$. 

Firstly, the $(k_x,k_z)=(0,0)$ Fourier mode of the streamwise velocity ($\overline{u}$)
evolves according to 
\begin{equation}\label{eq:eq_motion_ubar}
\pderiv{\overline{u}}{t} + \pderiv{}{y}\overline{uv} 
+ \epsilon_L \Big(\overline{uu} + \pderiv{}{y} (y\, \overline{uu})\Big) - \nu \frac{\partial^2 \overline{u}}{\partial y^2} 
=  f-\frac{dP}{dx},
\end{equation}
which is simply the horizontal average of \eqref{eq:mom_post_discard_conservative} (for $i=1$) with additional 
forcing terms. $f$ is the fringe-region forcing whose explicit form is detailed below, while 
$dP/dx$ is the constant pressure gradient driving the flow in the near-wall region. Equation 
\eqref{eq:eq_motion_ubar} is augmented with the no-slip condition at $y=0$ and the homogeneous 
Neumann condition $\partial_y \overline{u}=0$
at $y=L_y$. Note the parameter $\epsilon_L$ here necessarily has units of $1/\textrm{length}$, which is 
emphasized with the subscript `$L$'. After detailing the forcing $f$, $\epsilon_L$ will be related back 
to the nondimensional parameter \eqref{eq:eps_defn}.


Taking the horizontal average of the slow-growth continuity equation \eqref{eq:continuity_sg_full} gives 
\begin{equation}\label{eq:mean_continuity}
\pderiv{\overline{v}}{y} + \epsilon_L \, \pderiv{}{y} (y \, \overline{u}) = 0,
\end{equation}
and the no-slip condition implies that $\overline{v} = -\epsilon_L y\, \overline{u}$, 
which is of course the analogue of the relation \eqref{eq:apriori_sg}.

In contrast to $\overline{u}$, the evolution equation for the mean spanwise 
velocity $\overline{w}$ is not given by the horizontal average of \eqref{eq:mom_post_discard_conservative}
for $i=3$. Instead, the $\bigoh{\epsilon}$ contributions involving $\overline{w}$ 
are neglected, as in \citet{Spalart:1988}. Using 
$$ \overline{u'w'} = \overline{uw} - \overline{u}\,\, \overline{w}, $$
$\overline{w}$ evolves as 
\begin{equation} \nonumber 
\pderiv{\overline{w}}{t} + \pderiv{}{y}\overline{vw} 
+ \epsilon_L \Big(\overline{u'w'} + \pderiv{}{y} (y\, \overline{u'w'})\Big) - \nu \frac{\partial^2 \overline{w}}{\partial y^2} =  0,
\end{equation}
with the no-slip condition and a homogeneous Neumann condition at $y=0$ and $y=L_y$, respectively. 
This can be justified by the fact that the mean spanwise velocity $W = 0$ in each of the large 
scale simulations listed in table \ref{table:dns_eps}; it does not grow. Moreover, numerical 
stability issues can arise if one includes the $\bigoh{\epsilon}$ contributions involving $\overline{w}$, 
since the laminar equation for $\overline{w}$
$$ \pderiv{\overline{w}}{t} -  \nu \frac{\partial^2 \overline{w}}{\partial y^2} = -\epsilon_L \overline{u}\, \overline{w}$$
can exhibit exponential growth. Indeed, if $\epsilon_L < 0$ (as is the case for each large-scale
simulation in table \ref{table:dns_eps}) and $\overline{u}$ is frozen in time, 
then $\overline{w}\sim \exp(-\epsilon_L \overline{u} t)$. 

The Fourier modes $(k_x,k_z) \ne (0,0)$ evolve as 
\begin{align}
\pderiv{u_i}{t} + \pderiv{}{x_j}(u_i u_j) + \epsilon_L \Big(u u_i &+ \pderiv{}{y} (y u u_i) \Big)   
 + \pderiv{p}{x_i} - \nu \frac{\partial^2 u_i}{\partial x_j \partial x_j} = 0 \label{eq:eq_motion_uprime}\\
&\pderiv{u_i}{x_i} = 0, \label{eq:fluc_continuity}
\end{align}
in which the $\bigoh{\epsilon}$ contribution to continuity is neglected, as in 
\citet{Spalart:1988}. This appears to be a sensible approximation for boundary layers with mild 
$\beta$ values; recall from figure \ref{fig:tbl_stress_balances} that the streamwise
evolution of the turbulent kinetic energy made a negligible contribution to the mean stress balances
of the KS-$\beta1$ and BVOS-$\beta1.7$ flows in the near-wall region. 
The slow-growth momentum equation \eqref{eq:eq_motion_uprime} is 
augmented with the no-slip condition $u_i=0$ at $y=0$ and the no-flux conditions 
\begin{equation} \nonumber 
v = 0, \,\,\, \pderiv{u}{y} = \pderiv{w}{y} = 0 \qquad    \textrm{at } y = L_y. 
\end{equation}
The model equations are solved numerically using the  
velocity-vorticity formulation of \citet{Kim:1987ub}, which is derived from 
\eqref{eq:eq_motion_uprime} and \eqref{eq:fluc_continuity}
in the usual way \citep{lee2015direct}.

With all the equations of motion determined, the details of the forcing function $f$ 
in the mean streamwise evolution equation \eqref{eq:eq_motion_ubar} can now be specified. 
Its role is to provide momentum that will be
transported to the near-wall region, and it is non-zero only in the fringe region 
$y > L_y/2$. It is constructed in such a way that, for a given set of values 
$dP/dx$ and $\epsilon_L$, the model's equilibrium wall shear stress equals unity. 

More specifically, from 
\eqref{eq:eq_motion_ubar}, the model's mean streamwise stress balance is
\begin{equation}\label{eq:nwp_mean_stress_balance}
\tau_w + \frac{dP}{dx} y = \tau_{\rm model}(y) + \int_0^y f\, ds,
\end{equation}
where, from the relation $V = -\epsilon_L y U$ (which follows from  \eqref{eq:mean_continuity}), 
the model stress $\tau_{\rm model}$ is 
\begin{equation}\label{eq:model_stress_defn}
\tau_{\rm model}(y) = \nu \pderiv{U}{y} - \chevron{u'v'} - \epsilon_L \left(
\int_0^y \big[ U^2  + \chevron{u'u'} \big] \, ds + y \chevron{u'u'} \right).
\end{equation}
The no-flux boundary conditions imply that at $y=L_y$, \eqref{eq:nwp_mean_stress_balance} becomes
\begin{equation} 
\int_{0}^{L_y} f\, dy = \tau_w + \frac{dP}{dx} L_y + \epsilon_L \left(
\int_0^{L_y} \big[ U^2  + \chevron{u'u'} \big] \, dy + L_y \chevron{u'u'}\Big\vert_{y=L_y} \right) .\nonumber
\end{equation}
If one then constrains $f$ to satisfy 
\begin{equation}\label{eq:f_constraint_mean}
\int_{0}^{L_y} f\, dy = 1 + \frac{dP}{dx} L_y + \epsilon_L \left(
\int_0^{L_y} \big[ U^2  + \chevron{u'u'} \big] \, ds + L_y \chevron{u'u'}\Big\vert_{y=L_y} \right),
\end{equation}
the desired result $\tau_w =1$ will follow. Assuming ergodicity, the identity 
$$ \chevron{A} = \chevron{\,\overline{A}\,} $$
is true for any field $A$. 
Hence, if at each point in time 
\begin{equation}\label{eq:f_pointwise_constraint}
\int_{0}^{L_y} f(y,t)\, dy = 1 + \frac{dP}{dx} L_y + \epsilon_L \left(
\int_0^{L_y} \overline{uu}(y,t)\, dy  + L_y\, \overline{u'u'}(L_y,t)  \right) 
\end{equation}
holds, then \eqref{eq:f_constraint_mean} will result at equilibrium. The functional
form of $f$ is described in section \ref{subsec:comp_params} below. If one additionally 
sets the kinematic viscosity $\nu = 1$, then at equilibrium the model is scaled in 
viscous units, and the parameter $\epsilon_L$ reduces to the nondimensional $\epsilon$
introduced in equation \eqref{eq:eps_defn}.

\subsection{Physical parameters}
\begin{figure}
  \begin{center}
	\includegraphics[width=1.0\textwidth]{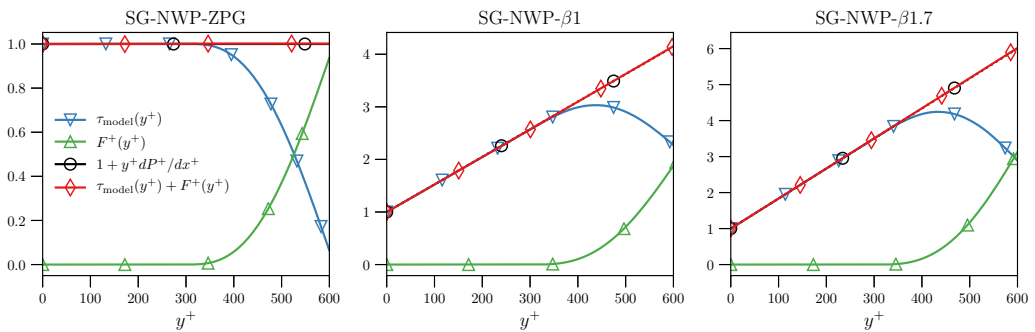}
  \end{center}
\caption{
For each case in table \ref{table:dns_eps}, the model stress $\tau^+_{\rm model}(y^+)$ 
(equation \eqref{eq:model_stress_defn}) agrees with the target stress profile $\tau_{\rm target}^+(y^+) = 1+y^+ dP^+/dx^+$
for $y^+\in [0,300]$, while in the fringe region $y^+\in [300,600]$, the primitive $F^+(y^+)$ of the 
forcing function $f$ supplies momentum flux so that $\tau_{\rm model}^+(y^+)+F^+(y^+)$ agrees with the
target stress throughout the entire domain $y^+\in[0,600]$.
} \label{fig:total_stress}
\end{figure}

\begin{table}
  \begin{center}
  \def~{\hphantom{0}}
    \begin{tabular}{c c c }
     Model case &  $dP^+/dx^+$  & $\epsilon$   \\
\hline
     NWP-ZPG           & $0$~ & $0$~          \\
     NWP-$\beta$1      & $5.503 \cdot 10^{-3}$~  & $0$~      \\
     NWP-$\beta$1.7    & $8.981 \cdot 10^{-3}$~  & $0$~ \\  
     SG-NWP-ZPG        &  $0$~                   & $-2.985\cdot 10^{-7}$~        \\
     SG-NWP-$\beta$1   & $5.503 \cdot 10^{-3}$~  & $-9.430\cdot 10^{-6}$~   \\
     SG-NWP-$\beta$1.7 & $8.981 \cdot 10^{-3}$~  & $-1.064\cdot 10^{-5}$~  
    \end{tabular}
    \caption{Imposed pressure gradient and slow-growth parameters for the model 
cases presented. Each value was chosen to match the corresponding one from the 
large-scale simulations listed in table \ref{table:dns_eps}. }
  \label{table:model_physical_params}
  \end{center}
\end{table}
Each slow-growth near-wall patch model case is parameterized by two inputs; 
they are the constant mean pressure gradient scaled in wall units $dP^+/dx^+$ and the asymptotic 
growth parameter $\epsilon$ given by \eqref{eq:eps_defn}. 
The values for the various model cases presented--two adverse-pressure-gradient cases
and one zero-pressure-gradient case--are shown in table \ref{table:model_physical_params}.
Note that for each pressure gradient value there is a model case both with ($\epsilon \ne 0$)
and without ($\epsilon =0$) growth effects included. 
Figure \ref{fig:total_stress} illustrates the statistically converged stress balances for 
the three slow-growth model cases. As the imposed pressure gradient $dP^+/dx^+$
increases, the total momentum transport in the near-wall region increases, as noted in 
section \ref{subsec:apg_tbl_discussion}. 

\subsection{Computational parameters and numerical implementation}\label{subsec:comp_params}
The remaining model parameters, consistent for all simulation cases, are 
summarized in table \ref{table:simulation_parameters} and are identical to 
those used in \citet{carney2020}. In particular the size of the 
rectangular domain $\Omega$ is taken to be $L_x^+ = L_z^+ = 1500$ and 
$L_y^+ = 600$, selected based on the spectral analysis of \citet{Lee:2019}. 
Their work suggests that, at least for the mild favorable-pressure-gradient 
cases considered, the contributions to the turbulent kinetic energy
from modes with wavelengths $\lambda^+ <1000$ are universal and $Re_{\tau}$ 
independent in the region $y^+ \lesssim 300$. 
Accordingly, $L_y^+$ is taken to be $2\cdot 300 =600$ to allow 
for a sufficiently large fringe region to mollify the effect of the non-physical 
computational boundary at $y=L_y$ (see figure \ref{fig:fringe_region}). 
The values $L_x^+ = L_z^+=1500$ were chosen because they were found to be
the smallest domain sizes capable of reproducing the universal small-scale 
turbulent kinetic energies identified in the channel flow simulations of \citet{Lee:2019} 
(see \S 3.3 in \cite{carney2020}). 
The majority of statistics reported in this work were from simulations with these
stream and spanwise dimensions; however, a range of values larger than 
1500 were also explored to assess the dependence of the model statistics on the 
choice of $L_x$ and $L_z$. 
In general, the statistics generated from these simulations exhibit little to no variation  
as the domain sizes grow. The streamwise and spanwise velocity variances
are two exceptions to this, however. As discussed more fully in section \ref{subsec:variances} below,
these statistics are known to be influenced by low-wavenumber, large-scale structures present 
in direct numerical simulations.  
As the NWP domain sizes grow, more of these large-scale structures are included, and
hence the stream/spanwise velocity variances continually change with increasing $L_x$ and $L_z$. 
The results are documented in the appendix \ref{sec:appendix_B}.

\begin{table}
  \begin{center}
  \def~{\hphantom{0}}
    \begin{tabular}{ c c c c c c c c}
       $L_x^+=L_z^+$  & $L_y^+$  & $N_x$ & $N_z$ & $N_y$ & $ \Delta x^+$ & $\Delta z^+$ & $\Delta y^+_{w}$     \\
       1500~           & 600~     & 120   & 256   & 192   &    12.5~     &  5.86~       &  0.002817~                 \\
    \end{tabular}
    \caption{Summary of simulation parameters consistent for all simulation cases; 
$N_x$ and $N_z$ refer to the number of Fourier modes, while $N_y$ is
the number of B-spline collocation points.
$\Delta x = L_x/N_x$ and similarly for $\Delta z$. 
 $\Delta y_w$ is the collocation point spacing at the wall.}
  \label{table:simulation_parameters}
  \end{center}
\end{table}

For a given selection of model parameters $(dP^+/dx^+, \epsilon)$, the forcing function 
$f$ responsible for providing momentum flux to the near-wall region is 
constrained at each timestep to satisfy \eqref{eq:f_pointwise_constraint}; otherwise, however, it 
is not uniquely specified. For the simulations reported here, $f$ is taken to 
be $f(y,t) = \psi(t) g(y)$, where $g$ is a piecewise cubic function
\begin{equation} \label{eq:g_defn}
g(y) =
\begin{cases} 
\begin{array}{rr}
4/L_y^4  \left(L_y-2y\right)^2 \left(5L_y-4y\right), &y \in [L_y/2,L_y] \\
0, &y \in [0,L_y/2]
\end{array}
\end{cases}
\end{equation}
which satisfies 
\begin{equation} \label{eq:g_constraint} 
 \int_{L_y/2}^{L_y}g(y)\, dy = 1 
\end{equation}
and $g(L_y/2)= g'(L_y/2) = g'(L_y) = 0$, 
so that the transition in forcing from the near-wall region to the fringe region is smooth.
In general, other function forms for $g$ are of course possible, and in particular a quadratic 
profile satisfying \eqref{eq:g_constraint} and $g(L_y/2)=g'(L_y/2)=0$ was also implemented with no discernible 
changes in the statistics in the near-wall region $y^+\in[0,300]$. 
The time-dependent function $\psi$ is defined as
\begin{equation} \label{eq:psi_constraint} 
\psi(t) = 1 + \frac{dP}{dx}L_y + \epsilon_L \left( \int_0^{L_y} \overline{uu}(y,t) \, dy + L_y \overline{u'u'} (L_y,t)\right),
\end{equation}
which is simply the right hand side of the forcing constraint \eqref{eq:f_pointwise_constraint}.
Together, \eqref{eq:g_constraint} and \eqref{eq:psi_constraint} ensure that the 
constraint \eqref{eq:f_pointwise_constraint} indeed holds.

As mentioned in \S 3.1 above, the model is solved numerically using the velocity-vorticity formulation
due to \citet{Kim:1987ub}. The numerical method is identical to the one employed by \cite{carney2020} 
and \citet{Lee:2015er}, that is, a Fourier-Galerkin method and a seventh-order B-spline collocation 
method for the stream/spanwise directions and wall-normal direction, respectively.  The equations of
motion are integrated in time with a low-storage, third-order Runge-Kutta (RK) method that treats 
diffusive terms implicitly and convective terms explicitly \citep{Spalart:1991wu}. Note that  
the forcing term $f(y,t)$ in the evolution equation \eqref{eq:eq_motion_ubar} for $\overline{u}$ is 
a nonlinear (and nonlocal) expression, and it is thus treated explicitly in the RK scheme, like 
the other nonlinear terms. 

The computational resolution in both time and space is chosen to be consistent with that of channel flow
DNS. The number of Fourier modes (and corresponding effective resolutions) used in each model simulation are
listed in table \ref{table:simulation_parameters}. They are comparable with, for example, the parameters
listen in table~1 in \citet{Lee:2015er}. 
Additionally, the collocation point spacing in the
wall-normal direction is similar to previous DNS studies; the total number of collocation points $N_y$
below $y^+ = 600$, as well as their distribution $\Delta y^+$ in the near-wall region, are taken to be equal to the
$Re_{\tau} = 1000$ case from \citet{Lee:2015er}. 
As in \citet{carney2020}, the model is implemented in a modified version of the PoongBack DNS 
code \citep{Lee:2013kv,Lee:2014ta}.

\subsection{Statistical convergence}
\begin{figure}
  \begin{center}
	\includegraphics[width=1.0\textwidth]{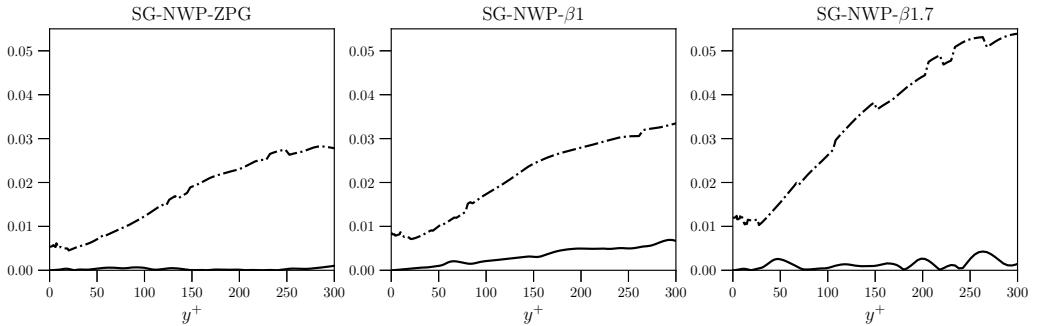}
  \end{center}
\caption{Statistical convergence for the slow-growth model cases
listed in table \ref{table:model_physical_params} -- solid lines: absolute error 
$ |\tau_{model}^+(y^+) - \tau_{target}^+(y^+)|$, where $\tau_{model}$ is defined
in equation \eqref{eq:model_stress_defn}, and $\tau_{target} = 
\tau_w + y\,  dP/dx$; dash-dotted lines: standard deviation of the estimated statistical 
error for $\tau_{model}^+$ in the region $y^+ \in [0,300]$. 
} \label{fig:stat_convergence}
\end{figure}

The method of \citet{Oliver:2014dh} is used to assess the uncertainty in the 
statistics reported due to sampling error. For each $(dP^+/dx^+, \epsilon)$ 
case listed in table \ref{table:model_physical_params}, the statistics are collected by 
averaging in time until the estimated statistical uncertainty in the 
total model stress profile $\tau_{\rm model}(y)$ (see equation \eqref{eq:model_stress_defn})
 is less than a few per cent. Figure \ref{fig:stat_convergence} shows
that the sampling error $|\tau^+_{\rm model}(y^+)-\tau_{\rm target}^+(y^+)|$ 
(where $\tau_{\rm target}(y) = \tau_w + y\, dP/dx$)
 is no larger than three per cent for each SG model case, 
and in particular the errors are smaller than the uncertainties; the errors
are similarly small for the $\epsilon =0$ cases (not displayed). 

\section{Numerical results}\label{sec:numerical_results}       
Based on the underlying scale separation assumptions, the slow-growth near-wall patch (SG-NWP) model 
can be interpreted in one of two complementary ways. In the first, 
one considers it to be a model for the small-scale near-wall turbulence
in a wall-bounded flow with (i) mean pressure gradient in wall
units $dP^+/dx^+$ and (ii) rate of change of the viscous length scale $\epsilon$ identical to 
those imposed in the model. In this case, one aspires to 
have the statistics from the model agree with those of the real flow at 
some fixed streamwise location. Ideally this match is exact for quantities 
that are insensitive to the unrepresented large-scale motions, while for
others the match holds after applying a high-pass filter, for example 
for the streamwise velocity variance, as in \citet{Lee:2019}.
This is the interpretation explored in the results reported here. 
In the second interpretation, the SG-NWP model represents the small-scale
near-wall turbulence in a region of real wall-bounded turbulent flow 
with local (i.e. scaled with the local wall shear stress) 
values of $dP^+/dx^+$ and $\epsilon$ the same as those imposed in the model. In this case, the SG-NWP
model is analogous to the universal signal of \citet{mathis_hutchins_marusic_2011}, representing
the process that is modulated by large-scale outer-layer fluctuations
in a real turbulent flow, and it is compatible with the quasi-steady, quasi-homogeneous 
description of the scale interactions in near-wall turbulence \citep{Zhang:2016,Chernyshenko:2021}.
It is this second interpretation where one could potentially employ the model to 
inform a pressure-gradient-dependent wall-model for an LES, for example. 

The statistics reported here were computed from SG-NWP model cases 
with three different pressure gradient values; for each value a 
separate simulation with $\epsilon=0$ was also conducted to help understand the
impact of asymptotic slow-growth effects.  
The model statistics are compared with the three large-scale simulation 
cases described at the beginning of \S 2.2: the zero-pressure-gradient case SJM-$\beta0$,
as well as the adverse-pressure-gradient cases KS-$\beta1$ and BVOS-$\beta1.7$. 
The corresponding model cases are thus referred to as SG-NWP-ZPG, SG-NWP-$\beta1$, 
and SG-NWP-$\beta1.7$ ($\epsilon\ne 0$), and NWP-ZPG, NWP-$\beta1$, and NWP-$\beta1.7$
($\epsilon=0$). The pressure gradients imposed in the model correspond to the local 
values scaled in viscous units of the large-scale simulations at the streamwise
location marked `$\times$' in figure~\ref{fig:utauvsx}, and the model's wall-normal 
statistical profiles are compared to those of the large-scale simulations at these
streamwise locations. All model cases are summarized in table~\ref{table:model_physical_params}.

\subsection{Mean velocity and shear stresses}\label{section:mean_vel_shear_stress}
\begin{figure}
  \begin{center}
	\includegraphics[width=1.0\textwidth]{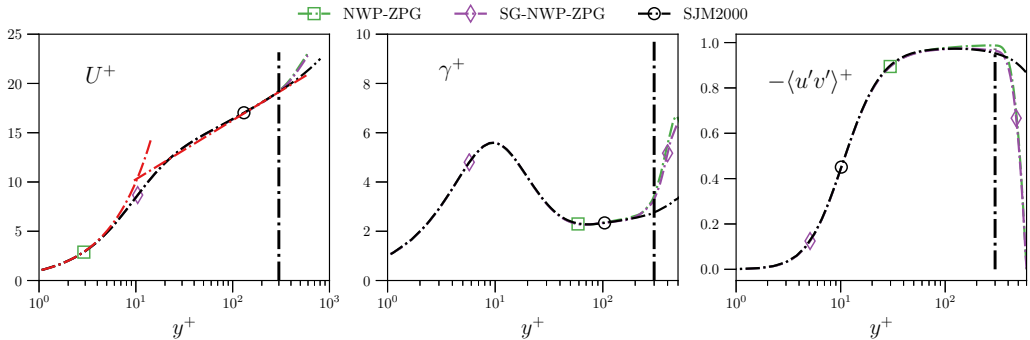}
  \end{center}
\caption{Mean velocity $U^+$ (left), 
indicator function $\gamma^+=y^+ \partial U^+/\partial y^+$ (middle), 
and Reynolds stress $\chevron{u'v'}^+$ (right) versus 
$y^+$ for the zero-pressure-gradient simulation cases. The
vertical, black dashed-dotted lines mark the beginning of the fringe
region $y^+=300$,  
and the law-of-the-wall 
$U^+ = y^+$ and $U^+ = (1/\kappa) \log(y^+) + B$ is also 
marked with a red dashed-dotted line (left), where $\kappa = 0.384$ and 
$B = 4.27$ \citep{Lee:2015er}.
}\label{fig:U_and_beta_zpg} 
\end{figure}

Because the Reynolds stress of a zero-pressure-gradient boundary layer is 
dominated by small-scale near-wall turbulent fluctuations and growth effects are
relatively insignificant, the mean velocity and Reynolds shear stress profiles 
from the NWP-ZPG model case are in excellent agreement with the corresponding 
DNS profiles \citep{carney2020}. The $\bigoh{\epsilon}$ terms included in the 
SG-NWP-ZPG case only enhance the agreement; in particular there is a modest 
improvement in the Reynolds stress profile for $y^+ \in [100,300]$
from the NWP-ZPG case that reduces the maximum relative error from 4\% to 1.3\%, 
as shown in figure \ref{fig:U_and_beta_zpg} (right). The slow-growth terms seem to not 
have an effect on the mean velocity $U^+$, as the NWP-ZPG and SG-NWP-ZPG profiles 
are nearly identical. In both cases, the relative 
error in $U^+$ is less than $0.6\%$ for $y^+ \in [0,300]$, and the error is similarly
small for the log-law indicator function $\gamma^+$
\begin{equation}
\gamma^+(y^+) = y^+\, \frac{\partial U^+}{\partial y^+}  \nonumber
\end{equation}
in the range $y^+\in[0,100]$. However, in both cases there is
mild disagreement of $\gamma$ in the range $y^+ \in [100,300]$. As expected, 
the profiles diverge for $y^+>300$; recall that $\chevron{u'v'}$ necessarily 
vanishes as a consequence of the $v=0$ condition posed (for Fourier modes $(k_x,k_z) \ne (0,0)$) 
at the upper computational boundary $y=L_y$. 


\begin{figure}
  \begin{center}
	\includegraphics[width=1.0\textwidth]{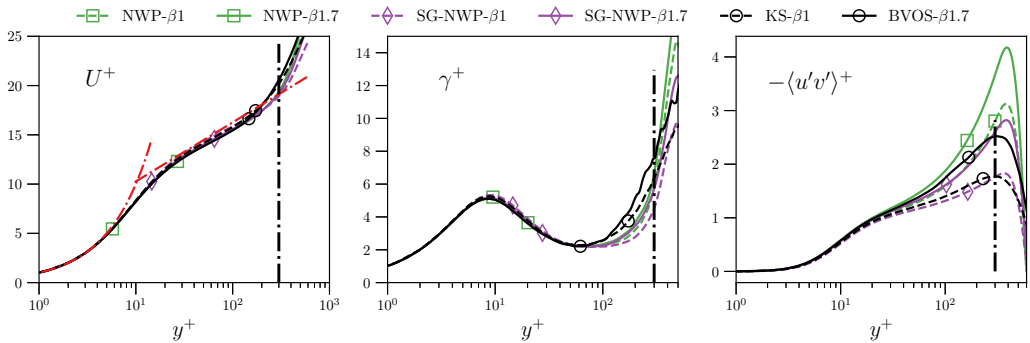}
  \end{center}
\caption{Mean velocity $U^+$ (left), 
indicator function $\gamma^+=y^+ \partial U^+/\partial y^+$ (middle), 
and Reynolds stress $\chevron{u'v'}^+$ (right) versus 
$y^+$ for the adverse-pressure-gradient simulation cases. The 
vertical, black dashed-dotted lines mark the beginning of the fringe
region $y^+=300$, and the law-of-the-wall 
$U^+ = y^+$ and $U^+ = (1/\kappa) \log(y^+) + B$ is also 
marked with a red dashed-dotted line (left), where $\kappa = 0.384$ and 
$B = 4.27$ \citep{Lee:2015er}.
}\label{fig:U_and_beta_apg} 
\end{figure}

Since growth effects fundamentally alter the balance of momentum transport
in the near-wall region for adverse-pressure gradient boundary layers (recall 
figure \ref{fig:tbl_stress_balances}), the model cases NWP-$\beta1$ 
and NWP-$\beta1.7$ that do not account for these effects are not expected to 
accurately reproduce the Reynolds 
shear stress profiles from the corresponding DNS and WRLES cases KS-$\beta1$
and BVOS-$\beta1.7$. Figure \ref{fig:U_and_beta_apg} indeed shows that 
this is the case; when $\epsilon =0$, the model severely overestimates the target
profiles for $y^+ \in [70,300]$. Including the $\bigoh{\epsilon}$ terms from the 
slow-growth analysis, however, leads to a significant improvement, as the 
SG-NWP-$\beta1$ and SG-NWP-$\beta1.7$ model cases both have a maximum 
error of about 9\% and 7\%, respectively, for $y^+\in[0,300]$. This remarkable 
improvement demonstrates that the SG-NWP model's $\bigoh{\epsilon}$ terms, 
as well as the forcing function $f$ in the fringe-region $300 \le y^+\le 600$,
 provide a good approximation of the momentum transport environment present in real,
spatially developing wall-bounded turbulent flows with mild adverse-pressure gradients. 

Similar to the zero-pressure-gradient case, the mean velocity and log-law
indicator profiles for the adverse-pressure-gradient model flows are 
essentially identical whether or not slow-growth effects are included. 
For all four adverse-pressure gradient model cases, the relative
error in the mean velocity profile $U^+$ is less than 6\% in the near-wall region 
 $y^+ \in [0,300]$. The error in the log-law indicator function $\gamma^+$ 
is similarly small for $y^+ \in [0,80]$, while each model case underpredicts
$\gamma$ for $y^+ \in [80, 300]$. However, this underprediction appears to be 
slightly worse for the SG-NWP cases than for those with $\epsilon=0$. 
From the mean stress balance \eqref{eq:nwp_mean_stress_balance}, for $\epsilon=0$, 
overprediction of the Reynolds shear stress implies underprediction of the mean viscous stress, 
however, the slow-growth terms present in \eqref{eq:nwp_mean_stress_balance}
appear to be responsible for the underprediction when $\epsilon \ne 0$, 
given that the Reynolds stress is more accurate. As discussed in section
\ref{subsec:scale_sep_assuptions}, it is expected that the accuracy of the model 
would increase in the region $y^+\in [80,300]$ if it were compared to a 
DNS/WRLES flow at the same pressure gradient $dP^+/dx^+$ and growth parameter $\epsilon$ but larger 
$Re_{\tau}$. 
%

\subsection{Velocity variances}\label{subsec:variances}

For a wall-bounded flow in a full size domain, the low-wavenumber 
contributions to the Reynolds stress represent the mean influences 
of the large-scale structures on the near-wall dynamics. It is well 
established that for channel and zero-pressure-gradient boundary 
layer flows, these low-wavenumber features of the near wall flow 
depend on $Re_{\tau}$
\citep{Hutchins:2007kd,Marusic:2010bb,LargescaleMotions:2017wl,Samie:2018,Lee:2019}.
Their contribution to the turbulent kinetic energy and their modulation
of the small-scale, high-wavenumber energy both increase with increasing $Re_{\tau}$. 
Experimental \citep{harun2013,sanmiguel2017,sanmiguel2020} and 
computational \citep{lee2017large,yoon2018,tanarro2020,pozuelo2022}
studies have demonstrated that a similar result holds for adverse-pressure-gradient
boundary layers; increasing the pressure gradient parameter $\beta$
leads to a significant enhancement of the large-scale energy, both in 
the outer layer and in the near-wall region. 

\begin{figure}
  \begin{center}
	\includegraphics[width=1.0\textwidth]{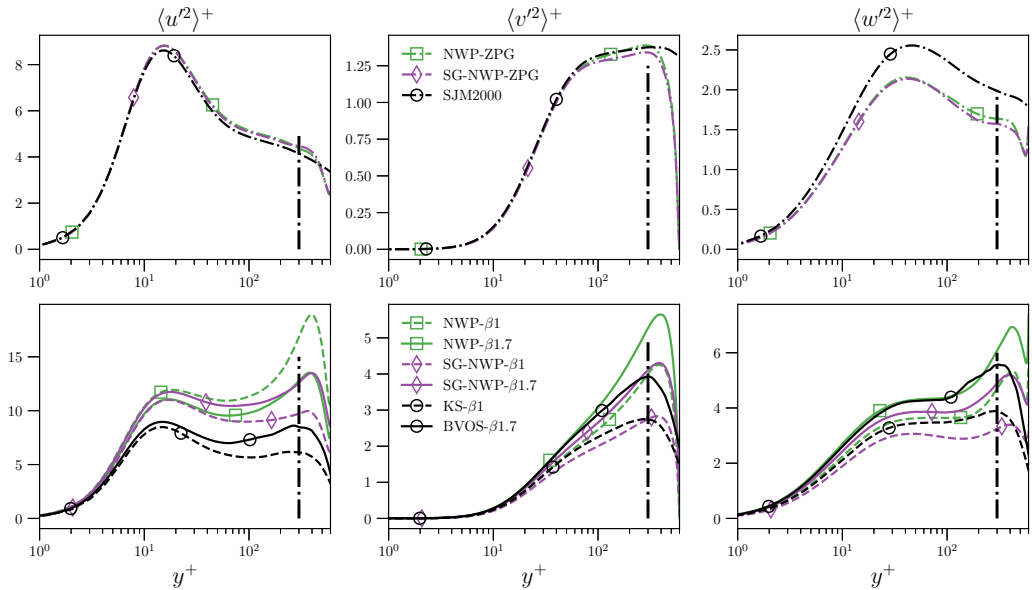}
  \end{center}
\caption{Velocity variances $\chevron{u'_{\alpha} u'_{\alpha}}$ versus $y^+$ for zero-pressure-gradient (top row) and 
adverse pressure gradient (bottom row) simulations.
The vertical, black dash-dotted lines mark the beginning of the fringe region $y^+=300$.  
} \label{fig:tke}
\end{figure}

The slow-growth near-wall patch model, by design,
cannot accurately represent these large-scale structures, as it 
instead seeks to isolate the dynamics of the near-wall small-scales
 associated with the autonomous cycle of \citet{Jimenez:1999wf} from 
their influence. 
Since the stream and spanwise velocity variances in both the zero and 
adverse pressure gradient boundary layers considered here 
are known to depend on low-wavenumber contributions, the 
corresponding model profiles are not expected to be accurate.
Figure \ref{fig:tke} shows this is indeed the case. 
Although both the NWP-ZPG and SG-NWP-ZPG $\chevron{u'u'}$ profiles 
appear to be in excellent agreement with the corresponding 
profile from \citet{sillero:2013}, this is a serendipitous coincidence.  
For an experiment or simulation at a larger $Re_{\tau}$, 
the streamwise velocity variance will grow 
(see figure~2 in \citet{Samie:2018}, for example), but it   
will be modeled by a similar SG-NWP flow; although $\epsilon$ 
may differ slightly, the pressure gradient will be the same. 
Similarly, at first glance, the model $\chevron{u'u'}$ profiles 
in the adverse-pressure-gradient cases (bottom row of figure \ref{fig:tke})
are seen to be significantly larger than their large-scale simulation counterparts. 
However, it should be noted that the 
KS-$\beta1$ and BVOS-$\beta1.7$ simulations are at relatively low
$Re_{\tau}$ (see table \ref{table:dns_eps}); at fixed $\beta$ 
values but larger $Re_{\tau}$, 
the profiles would be larger \citep{sanmiguel2020,pozuelo2022}. 

In contrast to the stream and spanwise velocity variances, the 
impact of large-scale motions on the wall-normal velocity 
variance for zero and adverse pressure gradient
boundary layers is less well-documented. For channel flows, 
\citet{Lee:2019} established that the $\chevron{v'v'}$ energy density 
in the near-wall region is concentrated primarily at wavelengths
less than $1000$ in viscous units, and hence the near-wall patch profiles 
from \citet{carney2020} were indeed in agreement with those from
large-scale simulations. Figure \ref{fig:tke} suggests a similar result
holds true for the near-wall region of adverse-pressure-gradient boundary 
layers. Although the agreement is not as good as the zero-pressure-gradient 
cases, the SG-NWP model $\chevron{v'v'}$ profiles compare 
reasonably well with the KS-$\beta1$ and BVOS-$\beta1.7$ data; the maximum 
relative error for $y^+ \in [0,300]$ is 14\% and 8.5\%, respectively. 
The NWP model without slow-growth effects, however, overpredicts the
adverse-pressure-gradient wall-normal velocity variance, similar to the Reynolds
shear stress profiles, illustrating the impact of the streamwise
development of the mean wall shear stress $\tau_w$ on the near-wall
turbulent kinetic energy. 

\subsection{Small-scale turbulent kinetic energy} \label{subsec:ss_tke}
Universal small-scale dynamics in the near-wall region associated with 
the autonomous cycle of \citet{Jimenez:1999wf} have been identified in 
channels \citep{Lee:2019} and zero-pressure-gradient boundary layers \citep{Samie:2018} 
by applying a high-pass filter to the energy spectral density. 
Increases in the near-wall, small-scale energy have also been reported 
in adverse-pressure-gradient boundary layers due to amplitude 
modulation effects \citep{harun2013,lee2017large,yoon2018}. \citet{sanmiguel2020} 
explicitly computed small-scale contributions to the near-wall turbulent 
kinetic energy. Using a cutoff wavelength of $\lambda_x^+ \approx 4300$,  
they found a collapse in the peak of the small-scale contribution to $\chevron{u'u'}^+$
at $y^+ \approx 15$ for a range of $\beta$ values from $0$ to approximately $2.2$.
In contrast, \citet{lee2017large} found that the small-scale contributions to the
streamwise velocity variance increased with the pressure gradient, however, this 
result was based on a much more restrictive high-pass filter; all contributions from 
spanwise wavelengths $\lambda_z^+ \gtrsim 180$ were filtered out. 
Using the same filter as \citet{carney2020} defined by \eqref{eq:filter_defn} below, 
the small-scale near-wall energy from SG-NWP flows is found to increase with increasing pressure
gradient, in agreement with the conclusion from \citet{lee2017large}.

In \citet{carney2020}, the near-wall patch model's high-pass filtered Reynolds
stress was directly compared to the filtered profiles of several channel flow DNS of 
\citep{Lee:2019}. After filtering out contributions from wavemodes that do not satisfy 
\begin{equation}\label{eq:filter_defn}
\min\{ |k_x|, |k_z| \} > k_{\rm cut} , 
\end{equation}
where $k_{\rm cut} = 2\pi/\lambda_{\rm cut}$ and $\lambda_{\rm cut}^+=1000$, the model's 
Reynolds stress profiles were in 
close agreement with the filtered DNS profiles, indicating the NWP model successfully 
reproduces the universal small-scale dynamics. Although the high-pass 
filtered Reynolds stresses of the present slow-growth NWP model cannot be directly 
compared to data from the literature, we report them here nonetheless to illustrate
the effect of pressure gradient and growth on the small-scale energies. 

\begin{figure}
  \begin{center}
	\includegraphics[width=1.0\textwidth]{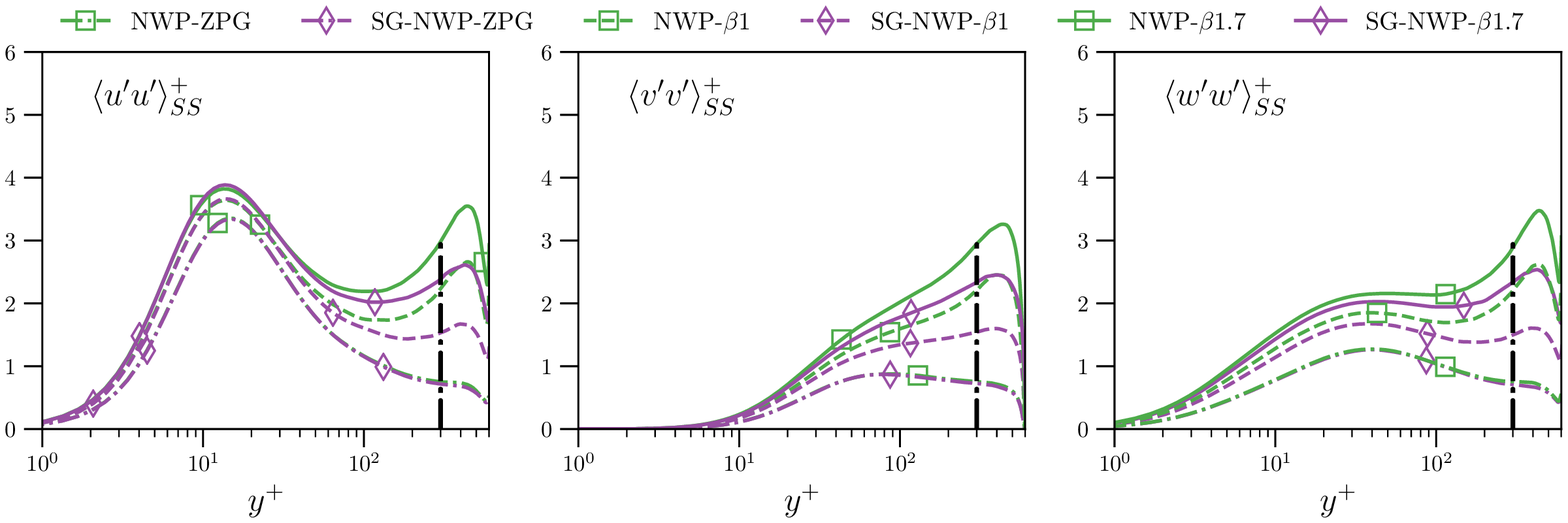}
  \end{center}
\caption{High-pass filtered velocity variances $\chevron{u'_{\alpha} u'_{\alpha}}_{SS}$ versus $y^+$ for each model case
listed in table \ref{table:model_physical_params}. 
The vertical, black dash-dotted lines mark the beginning of the fringe region $y^+=300$.  
} \label{fig:ss_tke}
\end{figure}

Let $\mathcal{K}$ denote the set of all wavenumbers included in a SG-NWP simulation, 
and let $k_{\rm} = 2\pi/\lambda_{\rm cut}$ with $\lambda^{+}_{\rm cut} = 1000$ as 
just mentioned. Define $\mathcal{K}_{SS}$ to be the subset of $\mathcal{K}$ with the
property that $(k_x,k_z) \in \mathcal{K}_{SS}$ if \eqref{eq:filter_defn} holds. 
If $E_{ij}$ denotes the Fourier transform of the two point correlation tensor 
$$
R_{ij} (r_x,y,r_z) = \chevron{u_i'(x + r_x, y, z + r_z) \, u_j'(x,y,z)}
$$
in the variables $r_x$ and $r_z$, then the small-scale energies shown in figure 
\ref{fig:ss_tke} are defined
to be 
\begin{equation}
\chevron{u_i'u_j'}_{SS} (y) = \sum_{(k_x,k_z) \in \mathcal{K}_{SS} } E_{ij} (k_x, y, k_z) . \nonumber
\end{equation}
It is clear that, for each component $\chevron{u'_{\alpha}u'_{\alpha}}$, the model's 
near-wall, small-scale energy increases as the pressure gradient $dP^+/dx^+$ increases, whether or not slow-growth
 effects are included. The agreement between simulations with and without 
$\bigoh{\epsilon}$ terms is particularly strong for the streamwise
small-scale energy in the region $y^+ \lesssim 50$, suggesting that the increase in small-scale,
near-wall energy in this region can be attributed soley to the adverse pressure-gradient effects, and not to growth effects. 

\subsection{Turbulent kinetic energy budget} \label{subsec:tke_budget}
The mean dynamics of the turbulent kinetic energy $k = u_i'u_i'/2$ are 
governed by the turbulent kinetic energy (TKE) budget equation 
\begin{equation}\label{eq:k_transport_eq_std}
\frac{D \chevron{k}}{Dt} 
= - \overbrace{\chevron{u_i'u_j'}\frac{\partial U_i}{\partial x_j}}^{\mathcal{P}_{k}}  
- \overbrace{\frac{\partial \chevron{ku_j'}}{\partial x_j}}^{T_{k}} 
+ \overbrace{\nu \frac{\partial^2 \chevron{k}}{\partial x_j\partial x_j}}^{D_{k}}
- \overbrace{\frac{\partial \chevron{p'u_j'}}{\partial x_j}}^{\Upsilon_{k}}
- \overbrace{\nu \chevron{\frac{\partial u_i'}{\partial x_j}\frac{\partial u_i'}{\partial x_j}}}^
{\varepsilon_{k}}. 
\end{equation}
For a wall-bounded turbulent flow that is homogeneous in the spanwise direction and slowly developing
in the streamwise direction, one can derive a `slow-growth' version of \eqref{eq:k_transport_eq_std} 
using the same multiscale asymptotic analysis outlined in section \ref{subsec:multiscale_analysis};   
namely, one transforms $x$ and $y$ according to \eqref{eq:coord_trans} and then inserts the 
scaling ansantz \eqref{eq:scaling_assumption} into the resulting equation. The resulting `slow-growth' 
TKE budget equation can be shown to be
\begin{align} 
\overbrace{2\epsilon \chevron{k}^+ U^+ }^{\textrm{SG Adv.}}
= &- \overbrace{\left( \chevron{u'v'}^+ \pderiv{U^+}{y^+} +  \epsilon \pderiv{}{y^+}(y^+U^+) \big[\chevron{u'u'}^+ - \chevron{v'v'}^+\big] 
\right)}^{\mathcal{P}^+_{k,\rm SG}}   \nonumber\\ 
&- \overbrace{\left( \pderiv{}{y^+}\chevron{kv'}^+ 
+ 3 \epsilon \chevron{ku'}^+ + \epsilon\, y^+ \pderiv{}{y^+} \chevron{ku'}^+   \right)}^{T^+_{k,\rm SG }}  \nonumber \\
& - \overbrace{\left(\pderiv{}{y^+} \chevron{p'v'}^+ 
+ 3\epsilon   \chevron{p'u'}^+ + \epsilon \, y^+ \pderiv{}{y^+} \chevron{p'u'}^+ \right)}^{\Upsilon^+_{k, \rm SG}} \nonumber \\
&+ \overbrace{\frac{\partial^2 }{(\partial y^+)^2}\chevron{k}^+}^{D^+_{k,\rm SG}} 
 - \overbrace{\chevron{\frac{\partial u_i'}{\partial x_j}\frac{\partial u_i'}{\partial x_j}}^+}^
{\varepsilon^+_{k,\rm SG}},  \label{eq:sg_tke_budget}
\end{align}
where the $\bigoh{\epsilon^2}$ contributions are neglected. 
If one alternatively starts with the slow-growth 
continuity \eqref{eq:continuity_sg_full} and momentum \eqref{eq:mom_post_freeze} equations from 
\S\ref{subsec:multiscale_analysis} and derives an equation governing the dynamics of $\chevron{u_i'u_i'}^+/2$
in the standard way, then, modulo $\bigoh{\epsilon^2}$ terms, nearly the same exact equation will result. 
The only difference is that, in this latter derivation, an extra $\bigoh{\epsilon}$ correction to the
dissipation $\varepsilon_k$ appears that is not present in \eqref{eq:sg_tke_budget}. The term is proportional 
to $y\chevron{u_i' \partial_x \partial_y u_i'}$ and arises from the $\bigoh{\epsilon}$ 
viscous term in \eqref{eq:mom_post_freeze}. 

Because the SG-NWP model momentum equation \eqref{eq:eq_motion_uprime} neglects some of the
$\bigoh{\epsilon}$ terms that arise from the asymptotic analysis in section 
\ref{subsec:multiscale_analysis}, there are some discrepancies between \eqref{eq:sg_tke_budget} 
and the model's TKE budget equation. In particular, there are discrepancies in the production 
$\mathcal{P}_{k,\rm SG}$, as well as the turbulent and pressure transport terms 
$T_{k,\rm SG}$ and $\Upsilon_{k,\rm SG}$. Although the errors introduced by these discrepancies 
are in each case relatively small, they are documented below when comparing the model's 
TKE budget to those of the large-scale simulations.

Firstly, since the mean advection  
$U^+ \partial_x^+ \chevron{k}^+ + V^+\partial_y^+ \chevron{k}^+ $ from the large-scale simulations
and the slow-growth advection 
$$2\epsilon \chevron{k}^+ U^+$$ 
from the NWP model cases have a maximum value no larger than $10^{-3}$, they are not plotted here. 

\begin{figure}
  \begin{center}
	\includegraphics[width=1.0\textwidth]{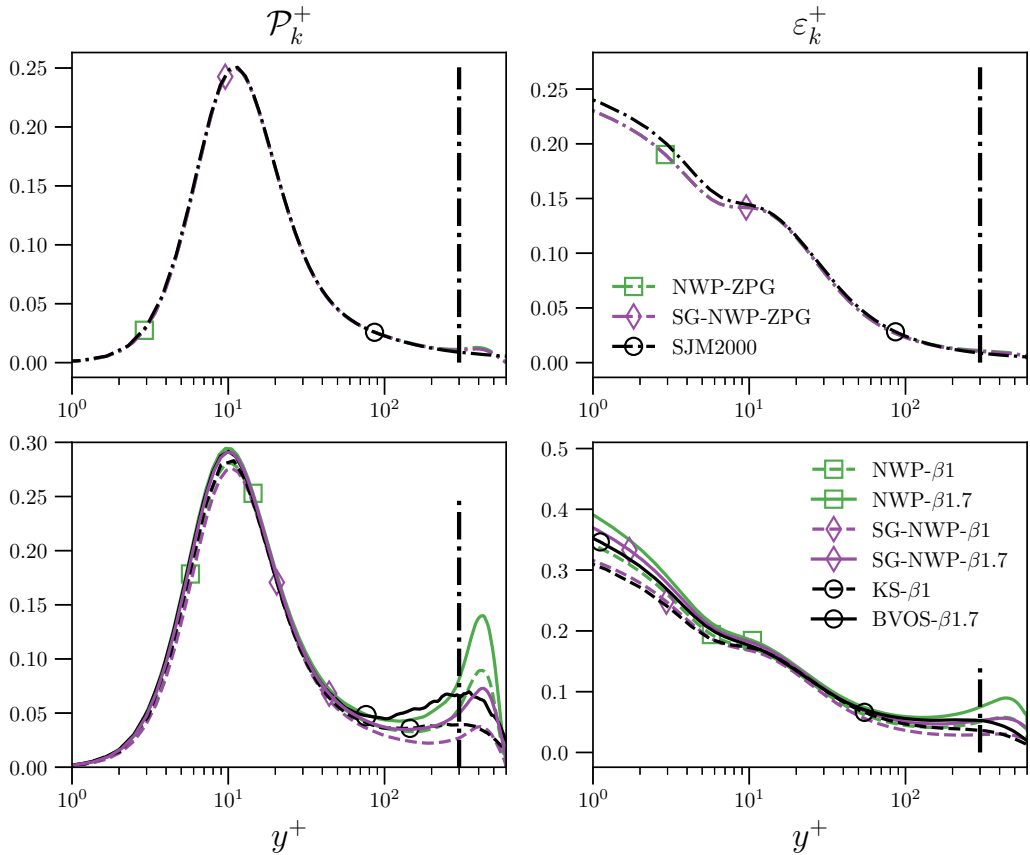}
  \end{center}
\caption{Left: production of TKE versus $y^+$, where the 
results from the large-scale simulations display $\mathcal{P}_k^+$ from 
\eqref{eq:k_transport_eq_std},
while the model results display the sum of $\mathcal{P}_{k,\rm SG}^+$ from equation 
\eqref{eq:sg_tke_budget} and \eqref{eq:k_production_spur}; right: 
dissipation of TKE versus $y^+$. For both quantities, the 
ZPG and APG flows are displayed in the top and bottom panels, respectively.
The vertical, black dash-dotted lines mark the beginning of the fringe region $y^+=300$.  
} \label{fig:prod_diss}
\end{figure}

For the production $\mathcal{P}_{k, \rm SG}$, the inconsistency between the continuity equations 
\eqref{eq:continuity_sg_full} and \eqref{eq:fluc_continuity} leads to some 
`spurious' $\bigoh{\epsilon}$ production terms for the SG-NWP model given by
\begin{equation}\label{eq:k_production_spur}
\mathcal{P}^+_{k,\rm spur} = \epsilon \left( U^+ \chevron{u'u'}^+ + \frac12 y^+ U^+ \pderiv{}{y^+}\chevron{u'u'}^+\right). 
\end{equation}
The left half of figure \ref{fig:prod_diss} displays $\mathcal{P}^+_k$ from \eqref{eq:k_transport_eq_std} for the 
large-scale simulation cases and the sum of $\mathcal{P}_{k, \rm SG}^+$ from \eqref{eq:sg_tke_budget} and the
spurious production \eqref{eq:k_production_spur}
for the SG-NWP flows. 
Noted that for the zero and adverse-pressure gradient flows considered in this work,
production is due almost exclusively to the product of $\partial U/\partial y$ and the 
Reynolds shear stress. For example, this term accounts for approximately~$97.5$\% of the total
production for the three large-scale flows from table \ref{table:dns_eps}. Similarly, the 
$\bigoh{\epsilon}$ terms account 
for only $0.2$\%, $5.7$\%, and $4.7$\% of the SG model's production $\mathcal{P}_{k, \rm SG}^+$ for 
$y^+\in [0,300]$
for SG-NWP-ZPG, SG-NWP-$\beta1$ and SG-NWP-$\beta1.7$, respectively, while the spurious production 
\eqref{eq:k_production_spur} is no larger than $0.1$\%, $3.7$\%, and $3.4$\% of $\mathcal{P}_{k,\rm SG}^+$ for these three
model cases. 


As expected, there is little difference between the SG-NWP-ZPG
and NWP-ZPG production profiles, since the mean velocity 
gradient and Reynold shear stress profiles for both  
model flows are nearly identical. Both are in excellent agreement with the 
SJM-$\beta0$ DNS case. For the adverse-pressure-gradient cases, 
the SG-NWP model's production $\mathcal{P}^+_{k, \rm SG} + \mathcal{P}^+_{k,\rm spur}$ is 
consistently smaller 
than the corresponding cases when $\epsilon =0$. At the near-wall 
production peak ($y^+\approx 12$), this results in better agreement
with the BVOS-$\beta1.7$ case for the SG-NWP model than the $\epsilon=0$
case, while for the KS-$\beta1$ case the agreement is slightly worse. 
In both cases, the slow-growth model underpredicts the 
production profiles from the large-scale simulations for $y^+ \gtrsim 60$, 
consistent with the results for the mean streamwise velocity gradient.

Unlike the production of turbulent kinetic energy, the dissipation
 $\varepsilon_{k} = \nu \chevron{\partial_j u_i'\partial_j u_i'}$
does not possess any additional slow-growth contributions, as previously
noted. The right half of 
figure \ref{fig:prod_diss} displays the dissipation profiles for the 
model and large-scale simulation cases. Although there is no discernible 
difference between the NWP model with
and without slow-growth effects in the zero-pressure-gradient case, they do impact the model dissipation 
profiles in the adverse-pressure-gradient cases. In particular they effect
a reduction in the maximum dissipation that occurs at the wall, resulting 
in better agreement with the large-scale simulation values. 


Similar to the dissipation, the viscous transport does not possess any $\bigoh{\epsilon}$ 
slow-growth contributions. However, 
there are indeed some $\bigoh{\epsilon}$ `spurious' contributions 
to the SG-NWP model's turbulent transport given by 
\begin{equation} \label{eq:k_turb_trans_spur}
T^+_{k,\rm spur} = \epsilon \left( \chevron{ku'}^+ + y^ + \chevron{k\pderiv{u'}{y}}^+ \right)
\end{equation}
that arise from omitting the $\bigoh{\epsilon}$ contribution to the fluctuating continuity 
equation \eqref{eq:continuity_sg_full}. The omission of both this term and the $\bigoh{\epsilon}$ 
pressure term in \eqref{eq:mom_post_freeze} from the SG-NWP model equations also leads to some
errors in the pressure transport. In particular, the model's pressure transport 
does not include any slow-growth contributions, in contrast to $\Upsilon_{k,\rm SG}$, so 
that in effect there is a spurious contribution given by 
\begin{equation}\label{eq:k_p_trans_spur}
\Upsilon^+_{k,\rm spur} = -\epsilon \left( 3 \chevron{p'u'}^+ + y^+ \pderiv{}{y^+} \chevron{p'u'}^+\right). 
\end{equation}

\begin{figure}
  \begin{center}
	\includegraphics[width=1.0\textwidth]{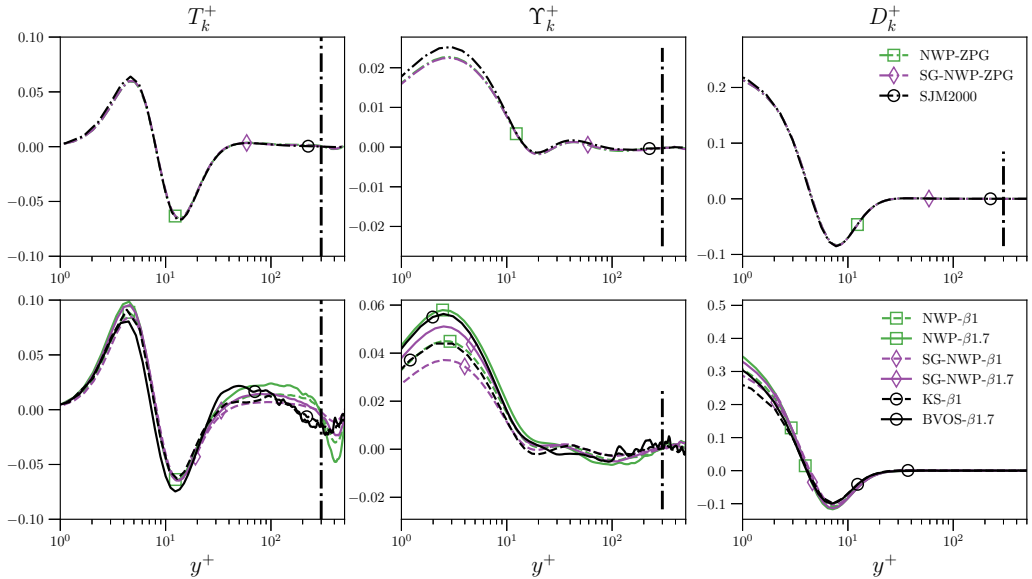}
  \end{center}
\caption{Turbulent (left), pressure (middle) and viscous (right) transport of TKE versus $y^+$.
The results from the large-scale simulations display $T_k^+$, $\Upsilon_k^+$ and $D_k^+$ from 
\eqref{eq:k_transport_eq_std}, respectively, while the model results display $T^+_{k,\rm SG}+T^+_{k,\rm spur}$, 
$\Upsilon^+_{k,\rm SG} + \Upsilon^+_{k,\rm spur}$, and $D_k^+$.  
The verticle, black dash-dotted lines mark the beginning of the fringe region $y^+=300$.  
} \label{fig:tke_trans}
\end{figure}

The SG-NWP turbulent $T^+_{k, \rm SG} + T^+_{k,\rm spur}$, pressure $\Upsilon^+_{k,\rm SG} + \Upsilon^+_{k,\rm spur}$,
and viscous $D^+_{k,\rm SG}$ transport profiles are displayed in figure \ref{fig:tke_trans} and show a reasonable agreement
to the corresponding profiles $T^+_k$, $\Upsilon^+_k$ and $D^+_k$ (as defined in \eqref{eq:k_transport_eq_std}) from the 
large-scale simulations. In particular, the SG-NWP model predicts a lower value of viscous transport at the wall than 
the corresponding model simulations with $\epsilon=0$, similar to the dissipation profiles. In contrast, the pressure
transport at the wall is notably less accurate when including slow-growth effects. The error is not due to the spurious
 pressure transport; the spurious terms generally contribute no more than $5$\% to the model's total turbulent and pressure 
transport profiles, similar to the TKE production. Since pressure is generally responsible for enforcing continuity, 
it is possible that the $\bigoh{\epsilon}$ error in the fluctuating SG continuity equation \eqref{eq:fluc_continuity} 
is responsible for the relatively large error. It would be useful to test this hypothesis in future work; 
a reformulation of the velocity-vorticity method due to \citet{Kim:1987ub} used here will be required to account for the 
fact that the velocity fluctuations are not divergence free.

\section{Discussion}\label{sec:discussion}
A primary goal of the slow-growth near-wall patch (SG-NWP) model is to understand 
how pressure gradients affect the small-scale motions in a wall-bounded turbulent flow. 
 The model is formulated by homogenizing a ``real'' patch of near-wall turbulence
and simulating the result in a restricted domain localized to the wall. The domain is 
periodic in the streamwise and spanwise directions, similar to the minimal flow unit 
simulations first conducted by \citet{Moin:1991}. 
Within the domain, the pressure gradient 
is assumed to be constant, while streamwise 
growth of the mean wall-shear stress that occurs in a turbulent
boundary layer, for example,  is accounted for asymptotically, 
similar to the related works by \citet{Spalart:1988} and \citet{topalian2017}. 
Far-field boundary conditions in the wall-normal direction are formulated by using
a fringe-region \citep{Colonius:2004} in which the mean flux of streamwise momentum in the wall-normal
direction is prescribed as a function of the pressure gradient and the asymptotic growth parameter. 

Overall, the model simulates \emph{only} the near-wall small-small motions. The dynamics of 
any turbulent structures larger than the NWP model domain size in the stream and/or spanwise 
direction are approximated only by the evolution of the Fourier modes of the velocity fields
with $k_x = 0$ and/or $k_z = 0$. By comparing the 
model statistics to those generated by a direct numerical simulation that resolves all scales 
of motion, the relative importance of the large-scale motions can be assessed. The 
relative importance of streamwise growth of the near-wall region can also be assessed
by comparing model cases that include the asymptotic terms (i.e.\ cases with $\epsilon \ne 0$) 
to those that do not (cases with $\epsilon =0$). The present investigation is hence in 
a similar spirit to the computational ``experiments'' that were recommended and conducted
by \cite{Jimenez:1999wf}.  

To this end, several conclusions can be drawn from the results presented in section \ref{sec:numerical_results}. 
Firstly, the mean velocity profile can be accurately computed for $y^+ \in [0,300]$ (to within about 6\%) without
resolving the large-scale structures. This accuracy is insensitive to whether asymptotic growth
effects are included or not. The Reynolds shear stress profiles can also be accurately computed (up to an 
error of about 9\%) without resolving the large-scale structures. 
The accuracy greatly deteriorates if growth effects are not accounted for
(especially for $y^+ \gtrsim 20$), however, highlighting the effect that the spatial development of the mean 
wall-shear has on the near-wall stress balance in adverse-pressure gradient flows. Similar conclusions
 can be drawn for the wall-normal velocity variance. In contrast, the SG-NWP model cannot accurately compute 
the stream and spanwise velocity variance statistics, as
the modulation and superposition effects known to be present
in channel flows, zero-pressure gradient TBLs, and adverse-pressure gradient TBLs are simply not
represented in the model.

Another important conclusion concerns the universality of small-scale motions in the
near-wall region. It is well established for channel flows and zero-pressure gradient 
boundary layers that the viscous-scaled turbulent kinetic energy (TKE) associated to small-scale motions
is independent of $Re_{\tau}$ \citep{Samie:2018,Lee:2019,carney2020,wang2021scaling}. The results here 
strongly suggest that for flows with adverse pressure gradients, small-scale universality in the near-wall
region should be pressure gradient dependent. 
The SG-NWP model's high-pass filtered, viscous-scaled turbulent kinetic energy (TKE) increased with 
the strength of the applied adverse-pressure gradient. 
It is natural to speculate that universality might be recovered if the small-scale TKE is instead scaled 
in so-called pressure-viscous units \cite{gungor2016scaling}, the nondimensionalization based on 
the kinematic viscosity $\nu$ and $u_{\rm pi}:= (\nu/\rho (dP/dx))^{1/3}$. This was not the case, however, 
in our numerical tests (not shown).
Consequently, the generalization to adverse-pressure gradient flows of the ``universal signal'' 
constructed by \citep{mathis_hutchins_marusic_2011} for zero-pressure gradient flows ought to depend on the APG strength.  

The current modeling approach of simulating only the small-scale motions is, in a sense, the 
reciprocal of what is done in large eddy simulations, where only the large scales are simulated, and 
the small-scales are approximated with a subgrid model. It is thus natural to try and link the SG-NWP 
model to an LES. As a quantitative model of near-wall turbulence, the SG-NWP model defines a two-parameter
family of near-wall turbulent flows, parameterized by the imposed pressure gradient $dP^+/dx^+$ and the growth 
parameter $\epsilon$. For such an application, one would invoke the scale-separation assumption discussed in 
section \ref{subsec:scale_sep_assuptions} and use SG-NWP flows matched to the local pressure gradient and 
growth parameter associated with the large-scale, outer-layer flow simulated by the LES. The coupling could 
be formulated sequentially (i.e.\ by precomputing a library of input and output responses) or concurrently
(i.e.\ ``on the fly'') \citep{abdulle:2012}. See  
\citet{sandham:2017,wang2021synthetic,elnahhas2021toward,chen2022locally} for examples of the latter.

Finally, as mentioned in the fundamental modeling assumptions discussed in section \ref{subsec:scale_sep_assuptions}, 
the SG-NWP modeling approach necessarily breaks down when the ``real'' flow to be modeled approaches separation, i.e.\ 
as $u_{\tau} \to 0$. The model is currently formulated to operate in viscous units; the forcing function 
$f$ in the fringe region ensures the model's wall shear stress at equilibrium equals unity. The strength 
of the forcing increases with $\epsilon$ and $dP^+/dx^+$, however, both of which blow up as $u_{\tau} \to 0$. 
Hence, an alternative scaling will be needed to improve on the current approach to model flows sufficiently 
close to separation. One starting point for 
future research may be to revisit the study by \citet{nickels2004inner} using data from recent large-scale
DNS and WRLES of separated flows, e.g.\ \citet{hickel2008implicit,gungor2016scaling}.

\section{Conclusions}\label{sec:conclusions}

The slow-growth (SG) model described here was formulated to extend 
the near-wall patch (NWP) representation of wall-turbulence presented 
in \citet{carney2020} to flows with non-negligible streamwise development 
of mean quantities. A primary objective is to provide a 
computationally accessible quantitative model of wall-turbulence 
for such situations, for example in boundary layers with adverse-pressure-gradients. 
Another is to characterize the extent to which
the dynamics of the small-scale motions, isolated from modulations
by large-scale structures, are responsible for observed characteristics
of near-wall turbulence. 
As in \cite{Spalart:1988} and \citet{topalian2017}, 
the model equations of motion are informed by asymptotic analysis 
of the Navier-Stokes (NS) equations; the fundamental assumption in 
the current setting is a separation between the viscous length scale 
and the length scale over which the mean wall shear stress evolves.

Because the SG-NWP model domain size scales in viscous units, the 
simulations require orders of magnitude fewer computational resources 
compared to large-scale DNS and wall-resolved LES. For example, 
the model's computational grid is approximately a factor of $5\, 700$ and 
$1000$ smaller than the DNS calculations of \citet{sillero:2013} and \citet{kitsios2017}, 
respectively, and a factor of about 144 smaller than the wall-resolved
LES of \citet{bobke2017}.

The SG-NWP model could hence be deployed as a vehicle for relatively
inexpensive numerical
`experiments' on near-wall dynamics influenced by pressure
gradients. 
For example, the model could be used to study the interactions of
the near-wall, small-scale dynamics with such complications as 
heat transfer, turbophoresis, or chemical reactions. 
The model should be particularly well suited for investigating the effects of 
 surface roughness on the near-wall region.

Finally, although the separation of scales on which the slow-growth
model is based is well-founded for a variety of practically relevant flows, 
it is important to keep in mind that it is still limited to a range
of pressure gradients that are not too large when scaled in viscous
units. Because the viscous length scale changes more rapidly with
stronger adverse pressure gradients, the asymptotic analysis on  
which the model is founded will no longer be valid.
In particular, this makes the model inadequate 
to describe near-wall flow near a point of boundary layer separation.

\section*{Acknowledgments}
The work presented here was supported 
by the National Science Foundation (Award No.\ 1904826), as well as by the Oden Institute for Computational 
Engineering and Sciences.
The research utilized the computing resources of the Texas Advanced Computing Center
(TACC) at The University of Texas at Austin. 
The authors are grateful 
to both Callum Atkinson and Ricardo Vineusa for sharing the statistics from 
large-scale adverse-pressure-gradient simulations.
The authors thank Gopal Yalla and Bj\"{o}rn Engquist
for insightful discussions, as well as Prakash Mohan for helpful suggestions
regarding time-integration of the equations of motion. 
Lastly, the authors thank the referees
for their many helpful suggestions that significantly improved the manuscript. 

\section*{Declaration of interests}
The authors report no conflicts of interest.

\appendix


\section{Effect of wall-parallel domain size on statistical quantities}
\label{sec:appendix_B}
A number of simulations were conducted at fixed values of $dP^+/dx^+$ and $\epsilon$ that 
correspond to the model case SG-NWP-$\beta1$ from table \ref{table:model_physical_params}
to assess the  sensitivity of the model's statistics to the size of the 
near-wall patch domain size in the stream and spanwise direction. In particular, 
simulations were conducted with $L_x^+ \times L_z^+ = 1500 \times 1500$ (as listed in table
\ref{table:simulation_parameters}), as well as with $1500\times 3000$, $3000\times 3000$, 
and $4500 \times 4500$. 
In each case, the number of Fourier modes $N_x$ and $N_z$ were increased proportionally 
to maintain the resolution $\Delta x^+$ and $\Delta z^+$ listed in table 
\ref{table:simulation_parameters}.

\begin{figure}
  \begin{center}
	\includegraphics[width=1.0\textwidth]{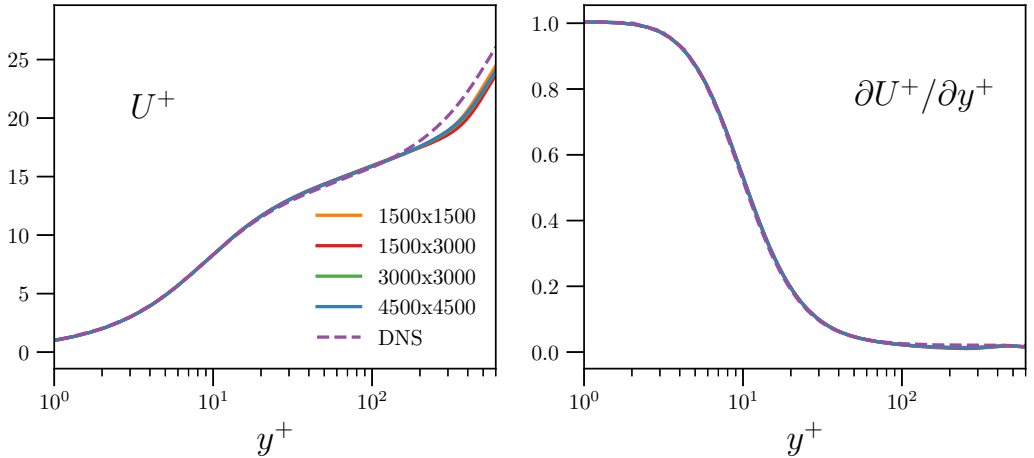}
  \end{center}
\caption{Mean velocity and mean velocity gradient profiles versus $y^+$ for the DNS adverse-pressure-gradient
simulation KS-$\beta1$ and SG-NWP-$\beta1$ model cases with various domain sizes $L_x^+ \times L_z^+$.} 
\label{fig:domain_size_compare_mean}
\end{figure}

\begin{figure}
  \begin{center}
	\includegraphics[width=1.0\textwidth]{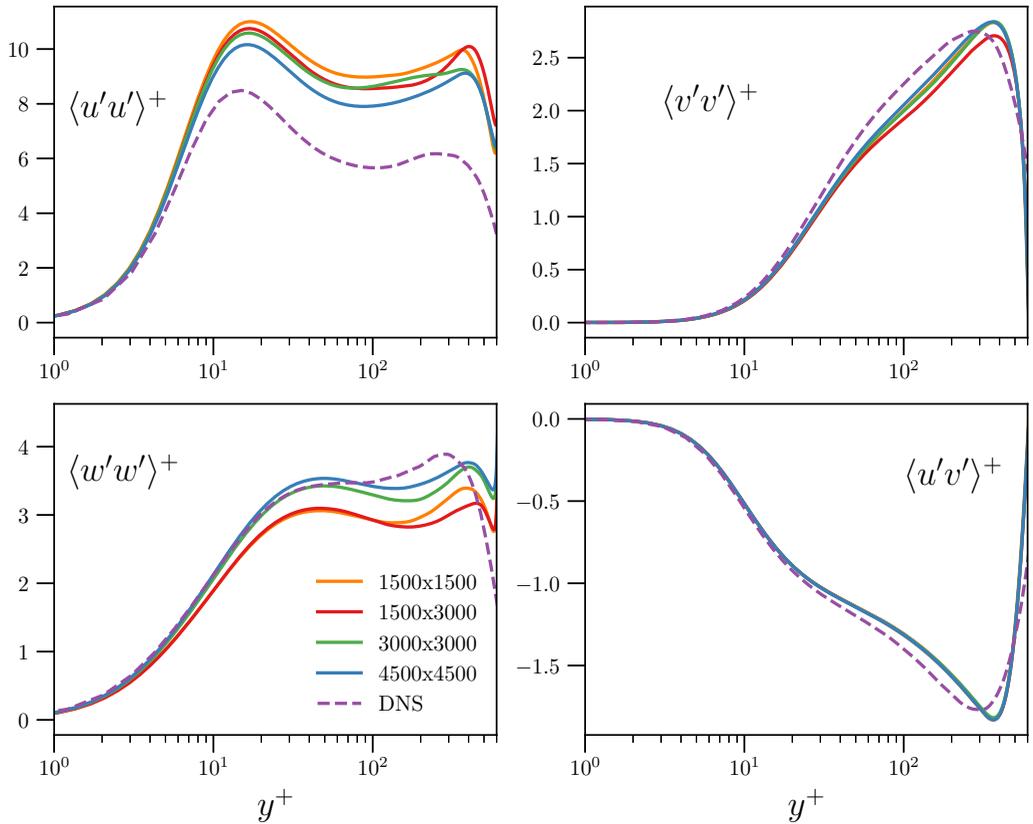}
  \end{center}
\caption{Reynolds stress profiles versus $y^+$ for the DNS adverse-pressure-gradient
simulation KS-$\beta1$ and SG-NWP-$\beta1$ model cases with various domain sizes $L_x^+ \times L_z^+$.} 
\label{fig:domain_size_compare_Re_stress}
\end{figure}

Figures \ref{fig:domain_size_compare_mean} and \ref{fig:domain_size_compare_Re_stress} show the 
mean velocity, the mean velocity derivative, and the Reynolds stress terms
for each domain size case, as well as the DNS profiles from \citet{kitsios2017}. 
The mean velocity, its derivative, and the Reynolds shear stress $\chevron{u'v'}$ all 
exhibit only slight variations as a function of the domain size. The wall-normal variances show a small increase 
with increasing $L_x^+$ and $L_z^+$, but the difference is less than a few percent. 
The turbulent kinetic energy budget terms (not shown) are also either unaffected, 
or show slight trends comparable to that of the wall-normal velocity variance.

In contrast, the streamwise velocity variances show a mild decrease with increasing domain size, while
the spanwise velocity variances show a nontrival increase.  
As mentioned in section \ref{subsec:comp_params}, 
the differences can be attributed to the fact that the NWP includes more large-scale structures
as $L_x^+$ and $L_z^+$ increase. At even larger domain sizes, the NWP size will begin 
to approach that of the full DNS; for example at $L_z^+ \ge 3000$, the NWP spanwise domain length is at least a 
third of that of the DNS, owing to the relatively low $Re_{\tau}$ at which it was 
conducted. In the limit as the NWP stream and spanwise domain sizes approach that of the DNS, 
it is reasonable to expect the NWP velocity variances to convergence to \emph{some} limiting profiles. 
However, these limits need not correspond to the DNS statistics, since the two flows are 
fundamentally different in multiple aspects. For example, they differ in the wall-normal domain size, the 
boundary conditions at $y=L_y$, and the use of a fringe region in the NWP. Another important difference is 
the treatment of boundary conditions in the stream and spanwise directions; the DNS uses a periodic 
conditions and recyling technique to account for boundary layer growth, while the NWP uses slow-growth 
asymptotics to account for streamwise growth in the near-wall region.  

\bibliographystyle{./jfm}

\bibliography{./references}

\end{document}